\documentclass[pre,twocolumn,amsmath,amssymb,showpacs,floatfix]{revtex4}
\usepackage{graphicx}

\newcommand {\revision}[1]{#1}
\newcommand{\bleq}{\ifpreprintsty \else
\end{multicols}\vspace*{-3.5ex}{\tiny \noindent\begin{tabular}[t]{c|}
\parbox{0.493\hsize}{~} \\ \hline \end{tabular}} \fi}
\newcommand{\eleq}{\ifpreprintsty \else
{\tiny\hspace*{\fill}\begin{tabular}[t]{|c}\hline
\parbox{0.49\hsize}{~} \\
\end{tabular}}\vspace*{-2.5ex}\begin{multicols}{2} \fi}

\newcommand{\bcols}{\ifpreprintsty\else\begin{multicols}{2}\fi}
\newcommand{\ecols}{\ifpreprintsty\else\end{multicols}\fi}
\newcommand{\te}{\theta}
\newcommand{\vzr}{v_z^0(r)}
\newcommand{\vzy}{v_z^0(y)}

\newcommand{\tnuly}{\tau ^0(y)}
\newcommand{\dvr}{\del v_r (r)}
\newcommand{\dvz}{\del v_z (r)}
\newcommand{\dvyp}{\del v_y (y)}
\newcommand{\dvzp}{\del v_z (y)}
\newcommand{\expo}{e^{i(kz-\omega t)}} 
\newcommand{\expor}{e^{i(kz-\omega_r t)}} 
\newcommand{\de}{\partial}
\newcommand{\om}{\omega}
\newcommand{\be}{\beta}
\newcommand{\eps}{\varepsilon}
\newcommand{\del}{\delta}
\newcommand{\tauaf}{\vec{\vec{\tau}} _{(1)}}
\newcommand{\vaf}{\vec{\nabla} \vec{v}}
\newcommand{\vafd}{(\vec{\nabla} \vec{ v}) ^{\dagger}}
\newcommand{\vu}{v^{\text{unp}}}

\begin{document}

\title{Weakly nonlinear subcritical instability
of visco-elastic Poiseuille flow}

\author{Bernard Meulenbroek$^1$}\thanks{Present address: CWI, Postbus
94079, 1090 GB Amsterdam, The Netherlands.}
\author{Cornelis Storm$^1$}\thanks{Present address: Institut-Curie,
  26 rue d'Ulm, 76005 Paris Cedex 05, France   }
\author{Alexander N. Morozov$^1$}
\author{ Wim van Saarloos$^{1,2}$} \thanks{Permanent address: Leiden}
\affiliation{$^1$Instituut--Lorentz, Universiteit Leiden, Postbus 9506, 2300 RA 
Leiden, The Netherlands\\ 
$^2$Laboratoire de Physique Statistique, Ecole Normale Sup\'erieure, 24 rue Lhomond,  
75231 Paris Cedex 05, France}

\date{\today} 

\begin{abstract} 
It is well known that the Poiseuille flow of a visco-elastic polymer fluid
between plates or through a tube is linearly stable in the zero
Reynolds number limit, although the stability is weak for large
Weissenberg numbers. In this paper we argue that recent experimental
and theoretical work on the instability of visco-elastic fluids in
Taylor-Couette cells and numerical work on channel flows suggest a scenario in which Poiseuille flow of
visco-elastic polymer fluids exhibits  a {\em nonlinear  ``subcritical'' instability} due to
normal stress effects, with a
threshold which decreases for increasing Weissenberg number. This
proposal is confirmed by an explicit weakly nonlinear stability
analysis for Poiseuille flow of an UCM  fluid. Our analysis
yields explicit predictions for the critical amplitude of
velocity perturbations beyond which the flow is nonlinearly
unstable, and for the wavelength of the mode whose critical amplitude
is smallest.  The nonlinear instability sets in quite abruptly at
Weissenberg numbers around 4 in the planar case and 
about 5.2 in the cylindrical case, so that for Weissenberg numbers
somewhat larger than these values  perturbations of
the order of a few percent in the wall shear stress suffice to make 
the flow unstable.   \revision{We have suggested elsewhere that this
nonlinear instability could be an important intrinsic route to  melt fracture
 and that preliminary  experiments are both qualitatively and
quantitatively in good agreement with these predictions.}

\end{abstract} 

\maketitle
%\bcols
%\begin{multicols}{2}

\section{Introduction} \label{secin}

\subsection{General motivation}
In this paper, we reconsider the classical topic of the stability of visco-elastic Poiseuille flow
in the zero Reynolds number limit. From a {\em weakly nonlinear
expansion}, we find that this flow is nonlinearly unstable for high
enough flow rates.

The first linear stability analysis of the flow of a
so-called Oldroyd-B fluid --- one of the simplest continuum models for a
visco-elastic polymeric fluid
with nonzero normal stress differences, characterized by a single relaxation
time $\lambda$ \cite{bird} ---  was already performed 
almost thirty years ago \cite{denn2}.  Since the subsequent careful
 linear stability analysis of Ho
and Denn \cite{denn1} it is generally accepted that 
 {\em Poiseuille flow of an Oldroyd-B fluid is linearly stable}, even
though the stability is weak for large values of the Weissenberg number, the
dimensionless 
quantity which measures the
strength of polymer relaxation effects. The definition used in this
paper \revision{for the case of cylindrical coordinates relevant for pipe flow is}
\begin{equation}
{\sf Wi} =  \left. \frac{\tau _{rr} - \tau_{zz}}{\tau_{rz}}\right|_{\text{wall}},\label{widef}
\end{equation}
where the term in the numerator is the normal stress difference of the
flow.
For the planar geometry \revision{ the index $r$ has to 
be replaced by $y$, with the $y$-axis taken  normal to the plates.} The term in the denominator of (\ref{widef})  is the shear stress at the wall. For an
Oldroyd-B or Upper Convected Maxwell (UCM)
fluid, the unperturbed flow field $v^{\text{unp}}$ is simply
parabolic, and we get
\begin{equation}
{\sf Wi} = 2 \lambda \left. \frac{\partial v^{\text{unp}}_z}{\partial
r}\right|_{\text{wall}} =   4 v_{\text{max}} \lambda/R  .
\end{equation} Here $v_{\text{max}}$ is the maximum velocity of the
unperturbed profile, 
$\lambda$ is the aforementioned relaxation time characterizing the
Oldroyd-B or UCM  fluid, and $R$ is the radius of the pipe.
For the planar case, R has to be replaced by $d$,
half the spacing between the plates.

\revision{It is well known that visco-elastic Poiseuille flow is linearly stable for
 well-established models like the UCM, Oldroyd-B \cite{denn1} and the Giesekus model \cite{keunings}.} However, there are good reasons to reconsider the {\em
nonlinear instability} of this flow configuration. First of all,
Atalik and Keunings \cite{keunings} have recently presented strong
numerical evidence tht visco-elastic Poiseuille flow is indeed {\em
nonlinearly unstable} for the UCM, Oldroyd-B and Giesekus model: when they injected the laminar flow field with a perturbation of
sufficiently large  amplitude,
instead of dying out (like one would expect for a strictly stable flow) the perturbation grew and saturated at a finite value, resulting in a finite amplitude oscillatory flow.  Such  behavior, in which the flow
is always linearly stable but nonlinearly unstable, is also called a
{\em subcritical instability}. A recent experiment
\cite{yesilata} also supports this nonlinear instability
scenario. Secondly, as we will discuss in detail in section
\ref{motivation} below, there are actually a lot of indirect
indications from well-established results on visco-elastic
Taylor-Couette flow that planar Couette or Poiseuille flow between
plates or in a tube might have a subcritical instability. Last but not
least is the following observation: if visco-elastic Poiseuille flow
is unstable, the nonlinear spatially and temporally oscillatory flow
pattern it will give rise to will inevitably result in distortions of
the flow after exiting the pipe or slit. In other words,
if visco-elastic Poiseuille flow is unstable for large enough ${\sf
Wi}$, it automatically leads to an intrinsic route to
``melt-fracture'' type phenomena, a generic name for the  fact that a polymer extrudate
normally develops strong undulations or irregularities beyond some critical flow
rate \cite{denn3,denn4,pahl}. It is well-established that there are various mechanisms that
can lead to such type of behavior (e.g., stick-slip behavior or exit
instabilities leading to sharkskin type patterns
\cite{denn3,denn4}). Most of these can be (partially) suppressed by
the proper choice of material or of the extruder shape and
coating. Our main conclusion (in line with \cite{keunings}) that
polymer Poiseuille flow is nonlinearly unstable for large enough ${\sf
Wi}$ implies that melt fracture type behavior is an unavoidable
consequence of a nonlinear instability in the extruder.  
\revision{In this paper, we focus on the subcritical instability of
  visco-elastic Poiseuille flow itself. Our arguments and experimental support
  for this scenario are summarized  in two
  recent letters \cite{meulenbroek,bonn1}, but the issue is complex
  and deserves further experimental and theoretical study.  We will 
  come back to it in the future.} 
%We simply refer for further discussion of this intrinsic route to meltfracture
%to two recent letters or ours. In the first
%\cite{meulenbroek}, the results which we derive here in detail were
%announced, while the second one \cite{bonn1} presents experimental
%evidence for the connection between the nonlinear flow instability and
% melt fracture type phenomena.

As stated, the calculations that we report in this paper give the
details of the explicit nonlinear amplitude expansion for the
stability of visco-elastic Poiseuille flow in the case of fixed
average pressure gradient  (we consider periodic modulations 
  of the pressure, so there are periodic pressure modulations but the
  average pressure drop remains unchanged).  We show explicitly that
for large values of ${\sf Wi}$ there  is  
indeed subcritical behavior, in other words that the flow exhibits a
weakly nonlinear instability. Moreover, the subcritical behavior sets in quite abruptly 
around a  value ${\sf Wi}_{\text c}$ of order 5 for flow in a
tube. This value is  consistent with the  value where melt fracture
type behavior is normally reported to set in 
\cite{denn3,denn4,pahl}, but  somewhat larger than the critical value
of 0.1 found numerically for for the related Oldroyd-B model with
viscosity contrast $10^{-3}$ and a Reynolds number of 0.1
\cite{keunings}. We will discuss the possible origins of these
discrepancies at the end of this paper. 

Since the nonlinear amplitude expansion that we will employ has been
used mostly in other fields of physics, 
\revision{and since it involves some unusual subtleties}, we present our results in some 
detail. For the benefit of the reader not interested in the
derivation, we  summarize our main results in section \ref{introsummary} of
this introduction. Before doing so,  we first 
discuss the relation of this work with the  Taylor-Couette problem.

\subsection{A nonlinear instability scenario motivated by the visco-elastic Taylor-Couette problem}\label{motivation}
Just over a decade ago,
Larson {\em et al.} \cite{larson} investigated the stability of 
the flow of a visco-elastic polymer solution in a Taylor-Couette  cell,
a cell consisting of two concentric rotating cylinders. They found that in
this system the
flow does
exhibit a well-defined linear instability for some value of the
Deborah number ${\sf De}$, which is analogous to the Weissenberg number. Their
calculations were done for the Oldroyd-B polymer model. In the
same paper the predictions were confirmed experimentally by a series
of measurements on a polymer solution which is well described by this
model. This work was later extended by Joo and Shaqfeh
\cite{shaqfeh0,shaqfeh1}, who considered the more general case of flow
in a curved channel. In the limit of boundary driven flow this reduces
to the Taylor-Couette case, while in the limit of static curved walls
the flow is 
driven by a pressure gradient (so-called Dean flow). In all  these
cases, as well as in the experimentally relevant cone and plate
geometry \cite{phantien1,phantien2}, the flow is linearly unstable at
large enough flow  velocities.  The
main conclusion from this line of research has therefore been that a linear
instability occurs  in visco-elastic 
fluids due to ``hoop stresses'' if the stream lines are curved 
\cite{shaqfeh2,pakdel}. This is confirmed by the observation that
the various stability calculations show that the instability threshold
goes to infinity if the curvature of the walls goes to zero. For the
Taylor-Couette system this is the limit in which the radius of the
cylinders goes to infinity; for the Dean flow problem investigated by
Joo and Shaqfeh \cite{shaqfeh0,shaqfeh1} this result is consistent
with the weak stability of Poiseuille flow between two parallel
planes.

In the last few years, Groisman and Steinberg \cite{steinberg1,steinberg2}
have experimentally investigated the
visco-elastic instability in the Taylor-Couette system in detail. They
find very good agreement with the theory for the 
onset of the  instability when the polymer flow velocity is increased;
however, an important  result of their careful study is that the
bifurcation at the critical Weissenberg/Deborah number is {\em
subcritical} or ``backward'' \cite{steinberg2,sureshkumar}.
%In some cases they see a hysteresis of
% the order of 50\%: once the instability has set in, they can follow
%the branch of  nontrivial visco-elastic finite-wavelength
%patterns to values of the velocity which are some 50\% below the value
%at onset. Although there had already been some theoretical hints that
%the bifurcation in the Taylor-Couette cell
%might be subcritical \cite{sureshkumar}, the
%experimental work by Groisman and Steinberg \cite{steinberg2} provides
%unequivocal evidence that this is  the case.

\begin{figure}
\begin{center}
  \includegraphics[width=0.9 \linewidth]{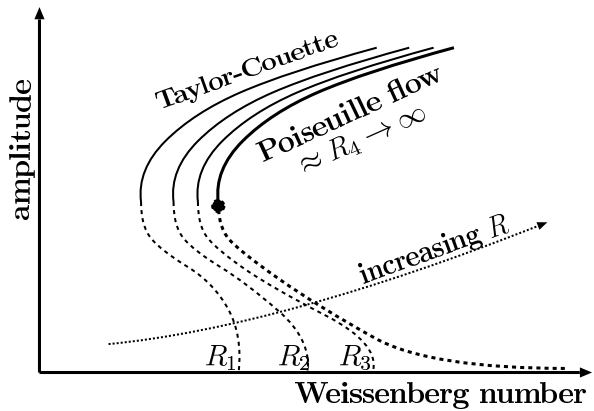}
\end{center}
\caption[]{Qualitative sketch of the subcritical behavior of the
visco-elastic instability in low Reynolds number polymer flow in a
Taylor-Couette cell, and of our proposal that  the bifurcation curve for the
case of Poiseuille flow should be thought of as the limit in which the radius 
in the Taylor-Couette cell goes to infinity. See text.
}\label{figconjecture}
\end{figure}

Groisman and Steinberg \cite{steinberg2}
have also presented intuitive arguments why the instability in
visco-elastic Taylor-Couette flow at low Reynolds numbers is
subcritical. In their arguments the fact that the flow is occurring
between curved walls is playing essentially no role: the curvature of
the walls is necessary to make the flow linearly unstable, but
the nonlinear positive feedback mechanism that in their picture  gives
rise to the subcritical behavior does not rely on the curvature. This
suggests the scenario sketched in Fig.~\ref{figconjecture}: The stable
nonlinear flow behavior, indicated in the figure by the full lines, is
quite independent of the precise radius or curvature of the cylinders,
assuming the ``gap'' (distance) between the cylinders to be constant.
The points where the curves touch the horizontal axis correspond  to
the linear instability threshold. These points therefore are strongly
dependent on the curvature of the cylinders: as indicated in the figure, 
the results of the theoretical analysis
\cite{shaqfeh0,shaqfeh1,shaqfeh2} imply that they shift to the right
as the radius $R$ of the inner cylinder increases, and that there is no
linear instability in the limit $ R\to \infty$.

The continuity found by Joo and Shaqfeh \cite{shaqfeh0,shaqfeh1} in
going from the Taylor-Couette flow to the case of Poiseuille-like  
Dean flow between  
fixed curved planes, as well as the intuitive arguments of Groisman
and Steinberg \cite{steinberg2} strongly suggest that the scenario for
Poiseuille-flow
through a channel will be very close to the one sketched in
Fig.~\ref{figconjecture} for the limit $R_4 \to \infty$: the
unperturbed flow is
linearly stable for any ${\sf Wi}$, but nonlinearly unstable as if there is
a subcritical bifurcation at ${\sf Wi}=\infty$. Moreover, just like the
rightmost curve for $R_4 \to \infty$  approaches the horizontal
axis rapidly for sufficiently large ${\sf Wi}$, implying that the threshold
for the nonlinear instability is small at large ${\sf Wi}$, we expect that
the threshold for the subcritical-like nonlinear instability of
visco-elastic Poiseuille flow is small for sufficiently large ${\sf Wi}$. 
The work done in this paper, summarized in the next subsection,
fully confirms this expectation.

It is of interest to note that the scenario we propose here for
visco-elastic fluids in the zero Reynolds number limit  has strong similarities
 for  large Reynold number planar Poiseuille flow of Newtonian fluids: planar
Poiseuille flow of ordinary fluids becomes linearly unstable at a Reynolds number of 
5772, but in practice the flow becomes unstable \revision{at much lower} Reynolds numbers of
 \cite{orszag,huerre,grossmann}.  \revision{Thus the
transition is  {\em subcritical}.  The scenario that  has emerged
 is that the nonlinear branch extends down to
${\sf Re}\approx 2500$  for two-dimensional perturbations and 
  to ${\sf Re} \approx 1000$ for three-dimensional flows, and in fact
already over thirty years ago} \cite{pekeris,hocking,herbert2,herbert1}, an amplitude 
expansion for planar Poiseuille flow in the spirit of ours was instrumental in showing  that the
instability at ${\sf Re}=5772$  is subcritical.  The
scenario we propose for visco-elastic flow is even closer to the one  that is found
for the transition to turbulence in Poiseuille flow of Newtonian
fluids  in a pipe: although the flow is also  linearly
stable for any ${\sf Re}$, the \revision{flow is nonlinearly unstable for ${\sf Re} \gtrsim 1000$ with a threshold which 
decreases  as ${\sf Re}^{-\gamma}$ for ${\sf Re }\to \infty$}
\cite{chapman}, very much like we suggest in
Fig.~\ref{figconjecture} with the dashed line labeled $R_4\to \infty$.
%(the scenario sketched in \cite{herbert2} for high Reynolds 
%number flow of Newtonian fluids is essentially the one sketched in
%Fig.~\ref{figconjecture} with the label $R_4 \to \infty$ for
%visco-elastic fluids).
The transition to turbulence of Newtonian fluids therefore shows that
whether or not there is a true  
linear instability is not very relevant in practice, and we believe that 
this is true for visco-elastic flows too.  
%Indeed, 
% conceptually, our work is very much in the spirit of that Pekeris and Shkoller
%\cite{pekeris}, who were the first to derive the first nonlinear 
% term in an amplitude expansion for the instability in planar
% Poiseuille flow of Newtonian fluids \cite{herbert1}.

Let us conclude this subsection with the following observation. As
noted in the preceding paragraphs, it is generally accepted that when the stream lines are
curved, visco-elastic flow becomes {\em linearly unstable} at sufficiently large flow
rates. In accord with this visco-elastic Poiseuille flow is linearly
stable since its stream lines are straight: infinitesimal perturbations must and do decay. However, following this same line of reasoning it is also very natural for visco-elastic Poiseuille flow
or planar Couette to be {\em nonlinearly unstable}: in the presence
of a {\em finite} perturbation, the stream lines {\em are} curved, and consequently we expect the flow to be
unstable to any finite perturbation for sufficiently large flow rates. This is the essence of the subcritical
instability scenario sketched above.

\subsection{Summary of our main result for  Poiseuille flow in the
UCM model}\label{introsummary}

In this paper we will consider the limit that  the Reynolds number
${\sf Re}$
\begin{equation}
{\sf Re} =  \frac{R v}{\eta / \rho}, 
\end{equation}
is negligible. Here
 $R$ is a characteristic length scale ($d$ for the planar case, the
radius $R$ for the case of a cylindrical tube),
$v$ a characteristic velocity, $\rho$ the density
and $\eta$ the viscosity of the fluid.
As ${\sf Re}$ measures the strength of the inertial
terms with respect to the viscous terms, in the limit ${\sf Re}\downarrow 0$
we can ignore the nonlinear convective terms in the momentum (Navier-Stokes)
equation, and make a quasistationary approximation in which the
temporal derivatives in the momentum equation are neglected.

As stated before, as the constitutive equation for the polymer fluid
we take the so-called UCM (Upper Convected Maxwell)  model
\cite{bird}, which expresses the 
stress tensor $\vec{\vec{\tau}}$ of the polymer fluid in terms of the
shear tensor $\vec{\nabla}\vec{v}$ through
\begin{equation} 
\vec{\vec{\tau }}+ \lambda \vec{\vec{\tau}}_{(1)} = - \eta
(\vec{\nabla} \vec{v} + (\vec{\nabla }  {\vec{v}})^{\dagger}),
\end{equation}
where  ``{\em the upper convected derivative}'' $\vec{\vec{\tau}}_{(1)}$ is given explicitly in
Eq. (\ref{deft1}) below, and where $\lambda$ is the  parameter with
the dimension of time that characterizes the UCM  model.

In this paper, we consider a perturbation of the velocity and stress
fields with single wavenumber $k$ along the direction of the flow, 
and with amplitude $A(t)$, i.e.
\begin{equation}
\mbox{perturbed fields} \propto A(t) e^{ikz} + c.c.
\end{equation}
where $c.c.$ means complex conjugate.
Then, in an expansion in powers of $A$, we determine to lowest
nonlinear order the equation for $A$,
\begin{equation} \label{Atriple}
\frac{{\rm d}A}{{\rm d}t} = -i\omega(k) A + c_3 |A|^2 A.
\end{equation}
To linear order in $A$ this equation simply reproduces the 
term  $i \omega(k)$ of the dispersion relation of a single mode
$e^{ikz-i\omega t}$; although we have to redo the linear stability
analysis in order to proceed to the nonlinear term, in principle this
term is already contained in the analysis by Ho and Denn
\cite{denn1}. In particular, since we know that every mode $k$ is
linearly stable, $\text{Im}\, \omega(k) <0$ for all $k$.
The essence of our analysis consists of calculating the coefficient
$c_3$ explicitly. In particular the real part of $c_3$ is of
importance: if the real part $\text{Re}\, c_3 <0$, then the nonlinear terms increase the
damping of the amplitude and the unperturbed state is, within this
approximation, not only linearly, but also nonlinearly stable. On the other hand, if $\text{Re}\, c_3
>0$, then the nonlinear term 
promotes the growth of the amplitude, and in particular amplitudes satisfying
\begin{equation}
|A| > A_{\text{c}}  = \sqrt{\frac{  \text{Im}\,  \omega(k)}{\text{Re}\, c_3}} \label{criticalamp}
\end{equation}
grow without bound. Hence, in this approximation $A_{\text{c}}$ defined above
constitutes the {\em critical amplitude of the perturbation beyond which the flow in
nonlinearly unstable}.

\begin{figure}
\begin{center}
  \includegraphics[width=0.9 \linewidth]{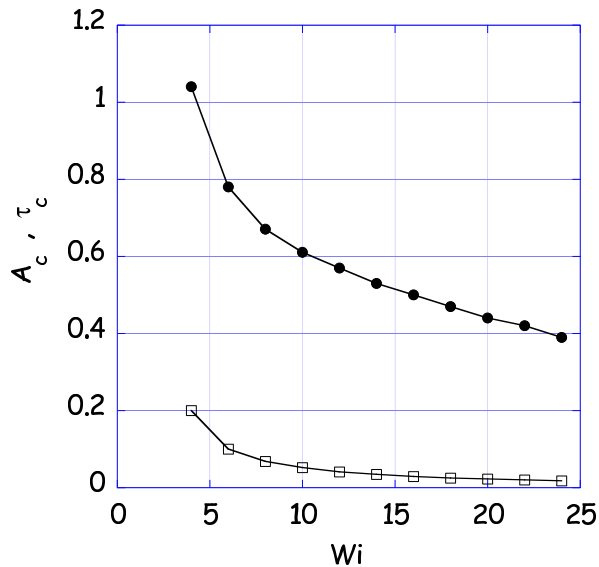}
\end{center}
\caption[]{The critical amplitude and the 
critical stress for the case of {\em planar }Poiseuille
flow of an UCM fluid, as determined from the weakly nonlinear
expansion in this paper.}\label{plaatje4} 
\end{figure}

\begin{figure}
\begin{center}
  \includegraphics[width=0.9 \linewidth]{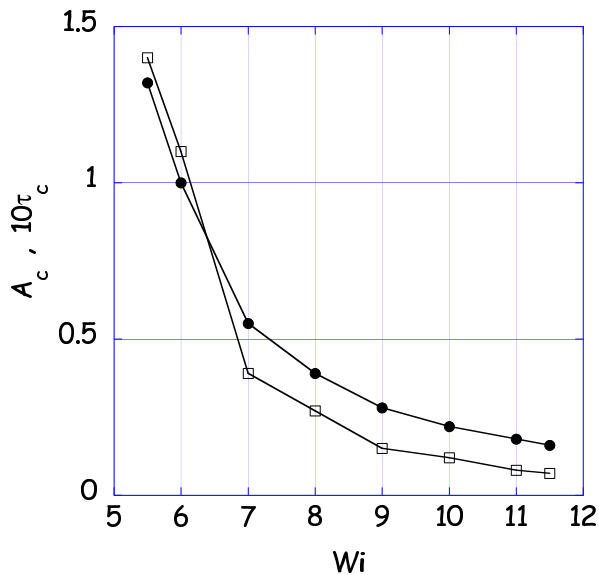}
\end{center}
\caption[]{The critical amplitude and the critical stress
for the {\em cylindrical } tube. Note that the critical values of
$\tau_{23}/ \tau_{23}^{\text{unp}}$ have been multiplied by 10 so as to be able to use the
same scale. }\label{plaatje5} 
\end{figure}

In our analysis, we do find that indeed for sufficiently large
Weissenberg number
$\text{Re}\, c_3>0$; we then determine the value of $k$ for which $A_{\text{c}}$ is
smallest, and take this as the critical amplitude for the nonlinear
flow instability. The value of $A_{\text{c}}$ obtained this way from our
analysis is plotted as a function of ${\sf Wi}$ in Fig.~\ref{plaatje4} for
the planar case  and in Fig.~\ref{plaatje5} for the cylindrical case. Our
normalization is such that $A$ is the ratio of the maximum perturbation in
the shear rate at the wall over one wavelength, divided by the unperturbed shear rate ,
\begin{equation}
|A| = \left. \frac{ \mbox{max} [\partial \, \delta v_z / \partial y ]}{\partial v^{\text{unp}}_z/
  \partial y} \right|_{\text{wall}} .\label{amplitude}
\end{equation}
As we
see from Figs.~\ref{plaatje4} and \ref{plaatje5} that the overall
behavior of the critical amplitude required to trigger the instability
is in accordance with the picture as suggested in Fig. 1: for  $R_4
\to \infty$, the planar limit, the threshold amplitude is expected to
get increasingly small as one increases ${\sf Wi}$, as indeed it does.

\begin{figure}
\begin{center}
  \includegraphics[width=0.9\linewidth]{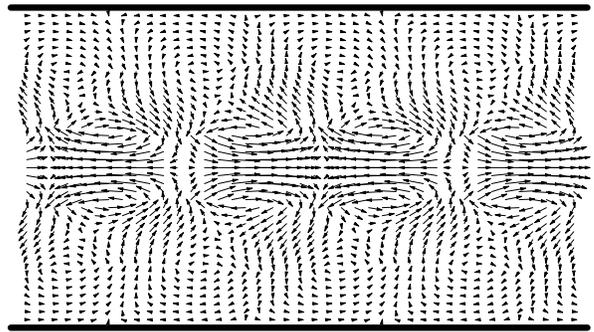}
\end{center}
\caption[]{Plot of the velocity field $\delta \vec{v}$ corresponding
to the linear eigenmode of the cylindrical geometry with 
wavelength $\lambda=1.7R$ at ${\sf Wi}=8$. The basic flow profile is in the
horizontal direction.}\label{velocityplotcyl} 
\end{figure}

In Fig.~\ref{velocityplotcyl} we show the velocity profile
$\delta \vec{v}$ corresponding with the linear eigenmode with
wavelength  $\lambda=1.7R$ in the planar case. This wavelength is
close to the one with the lowest instability threshold. The roll-type
structure of the flow-profile (which we also find in the planar case,
see section III) has very much the same structure as the
one in the Taylor-Couette cell that  according to the arguments
of Groisman and Steinberg \cite{steinberg2} underlies   the
subcritical instability in that case --- this confirms that
essentially the same mechanism is responsible for the subcritical
instability in Poiseuille flow.

An  important feature of our results which is of practical importance is
that the nonlinear instability sets in quite abruptly: below some
critical value  ${\sf Wi}_{\text{c}}$ of the Weissenberg number, the flow in
the cylindrical case is  within
our approximation also
nonlinearly stable as $\text{Re}\, c_3 <0$, while above ${\sf Wi}_{\text{c}}$ the critical
amplitude, especially the
critical amplitude for the shear stress at the wall, drops rapidly to a
small value (as we shall see, the case of planar Poiseuille flow is
slightly different). We can make this more precise as follows. In the
general 
scenario, the nonlinear branch where the flow-profile is nontrivial
and of the form sketched in Fig.~\ref{velocityplotcyl}, ends at a
so-called saddle-node bifurcation point --- this is the point marked
with a dot  in 
Fig.~\ref{figconjecture} where the unstable dashed branch and the stable
branch, indicated with a full line,  meet. For control parameters below
the value corresponding to the saddle-node bifurcation, the nontrivial 
flow pattern is not dynamically stable anymore. Now, as we extend our
expansion only up to the first nontrivial order in the amplitude, we
are unable, from our expansion, to precisely  locate the saddle-node bifurcation
point. However, it is reasonable to associate the approximate location 
with the point where $\text{Re}\, c_3=0$. In this approximation, our estimate
for the saddle-node bifurcation, and hence for the possible onset of
the modulations in the flow profile and hence in the distortions of
the extrudate,  is ${\sf Wi}^{\text{cyl}}_{\text{c}} \approx
5.2$. Note also that  the threshold value for the nonlinear
instability drops quite sharply when ${\sf Wi}$ increases beyond this value: 
Assuming even
careful experiments can not avoid perturbations of the order of a few
percent, we would expect to actually see the nonlinear state at ${\sf Wi}$
values somewhat larger than ${\sf Wi}^{\text{cyl}}_{\text{c}}\approx 5.2$  in the
cylindrical tube. This is quite consistent with  the Weissenberg value
${\sf Wi}_{\text c}$ where according to the literature
\cite{denn3,denn4,pahl} extrudates typically exhibit undulations and
deformations.  In the planar case the value of ${\sf Wi}^{\text{pl}}_{\text{c}}$ is somewhat
less sharply defined, according to our results, but for all practical
purposes it appears to be 
somewhat smaller than ${\sf Wi}^{\text{cyl}}_{\text{c}}$. We will discuss the competition between the
unperturbed laminar flow profile and the nonlinear profile further in
the concluding section \ref{secdisc}.

A few remarks concerning Figs.~\ref{plaatje4} and \ref{plaatje5}, which constitute the
main result of our analysis, are in order:\\
\hspace*{4mm} {\em (i)} Of course, our analysis is only based on an
expansion to lowest nontrivial order in $A$; one may therefore wonder
how much higher order nonlinear terms affect the result for
$A_{\text{c}}$. Clearly, for small ${\sf Wi}$ where $A_{\text{c}}$ becomes of order unity, our
results should {\em not} be trusted quantitatively: higher order
(quintic) terms will change the answer (for the planare case they may
even give rise to
the
existence of a true saddle-node bifurcation point). However, for large
enough ${\sf Wi}$, our results {\em can} be 
trusted quantitatively. This is because the linear damping term
$\text{Im}\,  \omega$ becomes arbitrarily small at large ${\sf Wi}$ (it varies roughly as
$\text{Im}\, \omega \sim 1/{\sf Wi}$), so that then the amplitude expansion for $A_{\text{c}}$
becomes nicely ordered. \\
\hspace*{4mm} {\em (ii)} Since we have only determined the cubic
nonlinearity in the equation, we can not say anything about the finite
amplitude at which the instability will saturate. It is in fact
questionable whether  the saturation amplitude can be determined from {\em
any}  perturbative method. One hint that this might be possible to a
reasonable approximation comes from the experiments of Bertola {\em et
al} \cite{bonn1,bonn2}, which indicate that the amplitude of the
perturbations of the extrudate just above ${\sf Wi}_{\text{c}}$ is rather small.\\
\hspace*{4mm} {\em (iii)} The scale on the vertical axis is {\em not}
arbitrary. In Figs.~\ref{plaatje4} and \ref{plaatje5} we have plotted the size of the shear rate
perturbation normalized to the shear rate at the wall in the unperturbed
case. Using the equation 
\begin{equation}
\left. \frac{\mbox{max} [\del \tau_{rz}] }{\tau^{\text{unp}}_{rz}}
\right|_{\text{wall}}
=|C(1)  A_{\text{c}}|, \label{stressamp}
\end{equation}
which holds both for the cylindrical case and the planar case\revision{(with $r$ replaced by $y$)}, with  $C(1)$
a numerical constant defined in Appendix 
~\ref{appendixa},  
we show in  Figs.~\ref{plaatje4} and \ref{plaatje5} the ratio of the perturbed shear stress
at the wall over the unperturbed shear stress at the wall, beyond which the
flow is unstable.  Note also the steep drop
of the curve for ${\sf Wi}$ just above ${\sf Wi}_{\text{c}}$: {\em for all practical purposes the
transition is quite sharp}. 
\\
\hspace*{4mm} {\em (iv)} Our analysis also yields an idea of the value of $k$ of
the mode with the smallest critical amplitude. For large enough ${\sf Wi}$
that our analysis can be trusted, we find typical values about twice
the diameter of the slit or the tube both in the planar case and in
the cylindrical case.  A precise comparison has to be based on
analyzing the frequency of the flow distortions measured at a fixed
position, however, since the flow distorts upon exiting the die --- see  section
\ref{secdisc}.

\subsection{Outline of the paper}
\revision{ We present our nonlinear analysis in some detail for various 
reasons. First of all, an amplitude analysis is normally used for problems with a true linear
instability; the use of the method for cases like this one without a true instability involves some
unusual nontrivial subtleties. Second, such an expansion makes use of left eigenvectors of
the linear operator, whose behavior and boundary conditions are quite intricate and worth
discussing. Third,  the analysis we introduce may be of relevant to
other visco-elastic flow problems as well.}

In order to perform our weakly nonlinear analysis, we first have to
reanalyze the linear stability problem.  The essential results are
summarized in section \ref{secla}. In order to facilitate comparison
with the earlier work by Denn and coworkers \cite{denn2,denn1}, we
write the equation for the stability eigenmodes  in terms of the stream
function, which  satisfies  a fourth order linear
differential equation. 
The  coefficients  that we have (re)derived for this differential
equation are given in appendices, and are the same as those given in
the appendix of \cite{denn2}. A feature not discussed in the earlier
work, however, is that there are various eigenmodes with different
symmetries in the vertical direction for the planar case.  In section
\ref{secna} we first rewrite the linear eigenvalue problem in a form which is
closer to the one usually found in derivations of an amplitude
expansion \cite{ch,walgraef}. This is then followed by a discussion of
the derivation of the cubic nonlinearity in the amplitude equation. A
somewhat special feature of our approach that should be kept in mind
 is that normally amplitude
expansions are used for an expansion around a true bifurcation point where a
particular mode loses stability. Here the relevant linear modes are
always weakly damped. This gives rise to some slight differences in
the formulation.

\section{Linear stability analysis} \label{secla}

After having formulated the problem for the planar case in the first
subsection, we will summarize the main  steps  of the linear stability
analysis
in the second subsection. Detailed expressions for the coefficients are
relegated to  to appendices. The numerical results for the dispersion
relation of the linear modes are presented  in the last subsection.

\subsection{Formulation of the problem}

We would like to investigate the linear stability
of polymeric flow between two plates, separate by a distance of $2d$,
and through a cylinder of radius $R$.
The direction orthogonal to the planes will 
be our $y$-direction, the direction of
the unperturbed flow will be the $z$-direction --- the advantage of
this convention is that both in the planar and in the cylindrical
geometry the mean flow is in the $z$-direction.
From now on we take the flow two-dimensional by putting
\begin{equation}
v_x=0
,~~~~~~
v_{y,z}=v_{y,z}(y,z),
\end{equation}
in the planar case and
\begin{equation}
v_{\te}=0,
,~~~~~~
v_{r,z}=v_{r,z}(r,z).
\end{equation}
in the cylindrical case.
The first step is the derivation of an expression for the unperturbed flow:
\begin{equation} \label{vun}
\vec{v}^{\text{unp}}= (0,0,\vu_z(y)),~~~~~~\vec{v}^{\text{unp}} = (0,0,\vu_z(r)).
\end{equation}
The next step is the perturbation of the unperturbed flow:
\begin{eqnarray} \label{lineqv}
\vec{v} &=&  \vec{v}^{\text{unp}} + (0,\dvyp,\dvzp) \expo, \\
\vec{v} &=& \vec{v}^{\text{unp}} + (\dvr,0,\dvz) ,\\
\tau _{ij} &=&  \tau _{ij} ^{\text{unp}} + \del \tau _{ij} \expo.
\end{eqnarray}
We will keep the terms linear in $\del \tau _{ij}$ and $\del v$
and write the $\del \tau _{ij}$ and $\del v_z$ in
terms of $\del v_y$ (planar) or $\del v_r$ (cylindrical) and its derivatives.
This will give us a fourth order equations for $\del v_y$ or $\del v_r$.
with boundary conditions.
Solving this equation yields the dispersion relation $\om (k)$; 
the imaginary part of $\om$ determines the stability of the flow.

\subsection{Equations and boundary conditions}

\subsubsection{The equations}
The flow is taken to be incompressible, so that  the conservation of
mass equation becomes the incompressibility condition 
\begin{equation} \label{diveq}
\vec{\nabla} \cdot \vec{v} = 0.
\end{equation}
Moreover, as was already explained in the introduction, since we are
interested in the zero 
Reynolds number limit,  the Navier-Stokes equation, expressing
conservation of momentum, 
reduces to the linear equation
\begin{equation} \label{defpro}
-\vec{\nabla} p  - \vec{\nabla } \cdot \vec{\vec{\tau }} = 0.
\end{equation}
We will take the curl of equation (\ref{defpro}) to 
eliminate the pressure. This leaves us with an equation
for the components of the stress tensor.
The UCM  model describes the stresses in
the polymeric fluid \cite{bird}:
\begin{equation} \label{OBeq}
{\vec{\vec{ \tau}}} + \lambda { \tauaf}  = -\eta(\vaf+\vafd),
\end{equation}
where the stress tensor $\tauaf$ is defined in the following way \cite{bird}
\begin{eqnarray} \label{deft1}
\tauaf & = & \frac{D {\vec{\vec{\tau}}}}{Dt} - (\vec{\nabla} \vec{v})^{\dagger} \cdot \vec{\vec{\tau}} - \vec{\vec{\tau}} \cdot (\vec{\nabla} \vec{v}), \\
\frac{D}{Dt}& = &\frac{\de}{\de t} + \vec{v} \cdot \vec{\nabla}.
\end{eqnarray}
Eq.~(\ref{OBeq}) illustrates that the UCM  model is characterized
by one time constant $\lambda$, which models the polymer relaxation
time. The ``upper convected derivative'' $\tauaf$ is the simplest
frame-independent formulation that implements this \cite{bird}. 

The UCM
constitutive equation  only models normal stress effects, no shear
thinning. This is illustrated by the well-known fact that upon
using the Ansatz (\ref{vun}) for $\vu$, we find
that the steady flow profile of a UCM fluid is still parabolic,
\begin{equation} \label{expvu}
\vu _z(y)=v_{\text{max}} \left[1-\left( \frac{y}{d}\right)^2\right],
~~~~~\mbox{planar case},
\end{equation}
where $v_{\text{max}}$ is the maximum velocity at the center line. Furthermore, for the stress
tensor of the basic parabolic profile, one finds for the planar case simply
\begin{eqnarray}
\tau^{\text{unp}}_{yz}&=& \tau^{\text{unp}}_{zy} = -\eta \frac{\partial \vu_z(y)}{\partial y},\\
\tau^{\text{unp}}_{zz}& =& -2 \eta \lambda \left( \frac{\partial
\vu_z(y)}{\partial y}\right)^2.
\end{eqnarray}
All other elements of $\tau^{\text{unp}}$ vanish. Note that $\tau_{zz}^{\text{unp}}$
is nonzero and proportional to the square of the shear --- this is the
normal stress effect.

The results in the cylindrical  case are:
\begin{equation}
\vu_z (r) = v_{\text{max}}\left[ 1-\left( \frac{r}{R}
\right)^2\right] ,
\end{equation}
and 
\begin{eqnarray}
\tau^{\text{unp}}_{rz}&=& \tau^{\text{unp}}_{zr} = -\eta \frac{\partial \vu_z(r)}{\partial r},\\
\tau^{\text{unp}}_{zz}& =& -2 \eta \lambda \left( \frac{\partial
\vu_z(r)}{\partial r}\right)^2.
\end{eqnarray}

The first step in a linear stability analysis is to
 linearize equation (\ref{OBeq}) to obtain expressions
for the $\del \tau$ which are linear in $\del v$. At this stage, it is
useful to introduce the stream function $\Phi$. Generally, the stream
function is introduced by writing (for the planar case)
\begin{equation}
v_y = \frac{\partial \Phi}{\partial z},~~~~~~v_z=-\frac{\partial
\Phi}{\partial y},
\end{equation}
so that the incompressibility condition (\ref{diveq}) is satisfied
automatically. For linear perturbations of the form (\ref{lineqv}) we
simply have $\Phi=\phi\; \expo$, so that
\begin{equation} \label{deffi}
\dvyp=ik\phi(y), \quad\quad\dvzp= -\frac{\de \phi(y)}{\de y}. 
\end{equation}
The stream function is slightly more complicated in the cylindrical case: 
\begin{equation} \label{deffic}
 \frac{ik}{r}\phi(r)=\dvr, \quad -\frac{1}{r}\frac{\de \phi(r)}{\de r}=\dvz.  
\end{equation}
It is also convenient to introduce dimensionless variables
\begin{eqnarray}
 \quad {\hat{\omega}}  &=& 
\frac{\omega d}{v_{\text{max}}},~~~~~~~~ \quad  \quad \hat k = kd, \nonumber \\
 \mbox{plane:}~~~~~\hat z & = &z/d ~~~~~~~~~\quad \hat t = t v_{\text{max}}/d, \label{dimensionlesspane}\\
\xi & = & \frac{y}{d}, \nonumber
\end{eqnarray}
for the planar die, and
\begin{eqnarray}
 \quad {\hat{\omega}}  &=& 
\frac{\omega R}{v_{\text{max}}},~~~~~~ \quad  \quad \hat k = kR, \nonumber \\
 \mbox{cylinder:}~~~ \hat z & = &z/R ~~~~~~~\quad \hat t = t v_{\text{max}}/R, \label{dimensionlesscyl}\\
\xi & = & \frac{r}{R}, \nonumber
\end{eqnarray}
for the cylindrical die.
 For notational simplicity, we will rename $\hat
k\Rightarrow k$, $\hat z \Rightarrow  z$ and $\hat t \Rightarrow t$
and $\hat{\omega} \Rightarrow \omega$; it just means that we 
have to remember that 
all lengths are henceforth measured in units of $d$ or $R$, inverse
lengths in units of $1/d$ or $1/R$, and times in unites of $d/v_{\text{max}}$ or $R/v_{\text{max}}$.
The  imaginary part of $\omega $ determines the
linear stability of the flow; the flow
is linearly stable if $\text{Im}\,  \omega  < 0$. 
 In order to facilitate comparison of our
explicit expressions for the linear equation, given in appendix
\ref{appendixa},   with the expressions of Rothenberger {\em et al.}
\cite{denn2}, we also introduce their dimensionless
number ${\sf S}$ \cite{denn2}, 
\begin{equation}
{\sf S}= \frac{{\sf Wi}}{2} = \frac{2 v_{\text{max}} \lambda}{d}.
\end{equation}

Using definition (\ref{deffi}) we obtain equations 
for the $\del \tau$ in terms of $\phi(y)$.
Equation (\ref{diveq}) is always satisfied because of
definition (\ref{deffi}), equation (\ref{defpro}) will
give us a fourth order equation for $\phi(y)$ of the following 
form:
\begin{equation} \label{deffieq}
\phi''''+\be_3\phi'''+\be_2\phi''+\be_1\phi'+\be_0\phi=0,
\end{equation}
where 
\begin{equation}
\be _i = \be_i(k,{\sf S},\omega; \xi).
\end{equation}
We used the symbolic manipulation
program Maple to find the explicit expressions for the $\be_i$:
these expressions are the same as those given by Rothenberger {\em et
al.} \cite{denn2} and can be found in the Appendix.

\subsubsection{Boundary and symmetry conditions, planar case}

The aim of the linear stability calculation is to determine $\omega(k)$ for fixed
 $k$ and ${\sf S}$, such that $\phi$
satisfies the usual stick boundary conditions:
\begin{equation}
v_y=0,\quad v_z=0 \quad \mbox{at} \quad \xi=\pm 1.
\end{equation}
For $\phi$ these boundary conditions translate into
\begin{equation} \label{bc1}
\phi(\xi=\pm 1) =0, \quad \phi'(\xi=\pm 1) =0 .
\end{equation}
Note that we have four boundary conditions, and a fourth order
equation. At first sight, one might think that therefore the equation
might have unique solutions for any $\omega (k)$. However, 
because of the linearity of the problem,  if we
find a solution $\phi(\xi)$, $C \phi(\xi)$ with $C$ an arbitrary
complex constant is a solution of equation (\ref{deffieq}) as well. 
We eliminate this arbitrary degree of freedom by setting
\begin{equation} \label{bc2}
\phi''(1)=1.
\end{equation} 
Since this eliminates two trivial degrees of freedom which do not
affect the solution, it is now clear that for a given $k$ and
${\sf S}$, one
or more unique branches of the 
complex quantity $\omega(k)$ will be fixed by the differential equation for
$\phi$. 

Because of the vertical symmetry of the problem, eigenfunctions
will either be asymmetric or symmetric. In the first
case the boundary conditions 
on the centerline are 
\begin{equation}
\mbox{asymmetric:}~~~v_y=0, \quad v_y''=0 \quad \mbox{at} \quad \xi=0,
\end{equation}
which implies for $\phi(y)$
\begin{equation} \label{bc3}
\mbox{asymmetric:}~~~\phi=0,\quad \phi''=0 \quad \mbox{at} \quad \xi=0 ,
\end{equation}
and in the second case the conditions are
\begin{equation}
\mbox{symmetric:}~~~v_y'=0, \quad v_y'''=0 \quad \mbox{at} \quad \xi=0,
\end{equation}
which implies
\begin{equation} \label{bc4}
\mbox{symmetric:}~~~\phi'=0,\quad \phi'''=0\quad \mbox{at} \quad \xi=0 .
\end{equation}
We will explore  the stability of both profiles 
below.

\subsubsection{Boundary conditions, cylindrical case}

We will take the usual stick \revision{boundary conditions
$v_r=0$, $v_z=0 $ at $\xi=\pm 1$.
In the center we require
$v_r = 0$, $v_z$ finite.
In terms of $\phi$ we have 
$\phi=0$, $\phi'=0 $ at $ \xi=1$,
and we choose $\phi''(1)=1$.
We have to be careful in the origin, because of
the $1/\xi$ terms in the equations for the $\be _i$.
The boundary conditions on $\vec{v}$ imply
$\phi =0$, $\phi'=0 $ at $ \xi =0$.
In order to avoid numerical problems, we use a polynomial
$\phi(\xi) =a_2 \xi^2 / 2+  a_3 \xi^3/6 $
for $\xi \in (0,0.01)$.
At $\xi =0.01$ we match both solutions by  imposing  equality of
$\phi$ and the first three derivatives.}

\subsubsection{Summary of the linear stability problem}\label{summary}

In summary, we have to solve equation (\ref{deffieq}) with the
condition  (\ref{bc2}) and the 
boundary conditions (\ref{bc1}) and (\ref{bc3}) for asymmetric modes 
or with the boundary conditions (\ref{bc1}) and (\ref{bc4}) in the
case of symmetric modes. 
This means that we have to find the two \revision{complex parameters}
$\phi'''(1)$, $ \omega (k)$,
such that the conditions on the centerline are satisfied.
In the cylindrical case we have to solve equation (\ref{deffieq}) with
the boundary conditions discussed above.
This means that we have to find the four complex parameters
$\phi'''(1)$, $\omega$, $a_2$ and $ a_3$
such that both solutions can be matched together.

\subsection{Numerical results}\label{seclanum}
\subsubsection{The planar case}
We have used a shooting program \cite{numrec} to find the right parameters
and to construct the eigenmodes corresponding to the
two classes of boundary conditions. We present the results below for the
range of $k$-values for which our program converged essentially to
arbitrary  precision.  For larger values of
$k$ the convergence becomes poorer, but since these larger values do
not appear to be relevant for the nonlinear analysis of section
\ref{secna}, we content ourselves with reporting simply the range where
sufficient precision could be reached.

\begin{figure}
\begin{center}
  \includegraphics[width=0.9\linewidth]{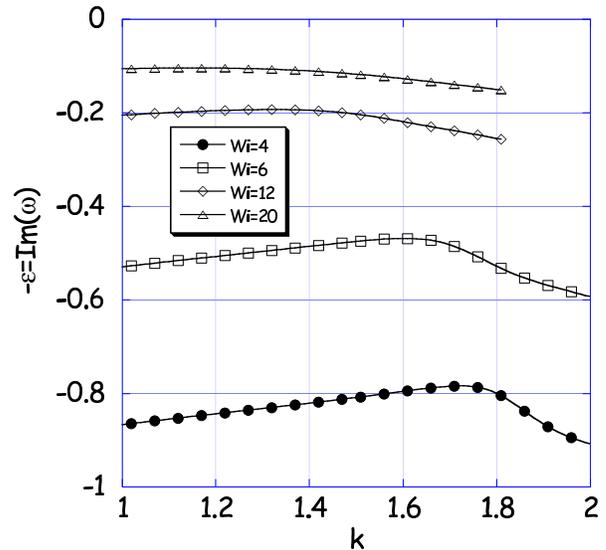}
\end{center}
\caption[]{Dispersion relation for $\text{Im}\, \omega=-\varepsilon$ of the weakest damped asymmetric mode
for  ${\sf Wi}=4, 6, 12$ and $20$ in the planar case.}\label{plaatje1}
\end{figure}

Fig. \ref{plaatje1} shows  the magnitude of the
imaginary part  of the 
eigenvalue $\omega$ of the {\em
asymmetric mode}  as a function
of $k$ and for various ${\sf Wi}$. 
For all values of ${\sf Wi}$, $ \text{Im}\, \omega  <0$. As the temporal behavior of the modes
is as $e^{- \varepsilon t}$, this confirms 
that the flow is  {\em linearly stable}. 
Especially for large  ${\sf Wi}$, $\varepsilon$ is very
small, which means that the flow is only weakly 
stable. This is the reason that we also introduce the dimensionless
temporal decay rate   $\varepsilon =- \text{Im}\, \omega$,  so that the positive
quantity $\varepsilon$ is a 
small quantity for sufficiently large ${\sf Wi}$. This is very important, as
the smallness of $\varepsilon$  will allow us to use an 
amplitude expansion in the next section: this expansion is based 
on an adiabatic approximation for the growth of the amplitude relative
to the intrinsic oscillation of the
waves with frequency $\text{Re}\, \omega$. Since the intrinsic 
frequency is of the order of unity in our dimensionless units, the
condition for the amplitude expansion to work is that 
$\varepsilon  \ll  1$.

\begin{figure}
\begin{center}
  \includegraphics[width=0.9\linewidth]{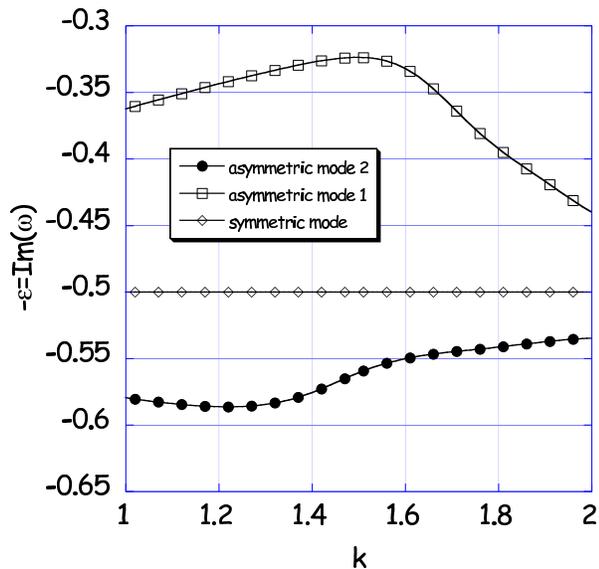}
\end{center}
\caption[]{The dispersion relation for the two different linear
asymmetric eigenmodes at ${\sf Wi}=8$ in the planar case, and the symmetric
mode whose damping rate  is given exactly by Eq.~(\ref{exact}).}\label{plaatje3}
\end{figure}

We found a second asymmetric mode close to the first
one reported in Fig. \ref{plaatje1}. This mode is
slightly more damped than the first one. The two asymmetric modes are
shown   in Fig.~\ref{plaatje3}.  The symmetric mode lies
between the two asymmetric modes, for, 
quite surprisingly, we have been able to show analytically that the
symmetric mode obeys 
\begin{equation}
\text{Im}\, \omega (k) = -\frac{2}{{\sf S}}  \quad \Longleftrightarrow \quad \eps =
\frac{2}{{\sf S}}=\frac{4}{{\sf Wi}}. \label{exact}
\end{equation}
so that for the the Weissenberg number ${\sf Wi}=8$ in Fig.~\ref{plaatje3}
the damping rate of the symmetric mode is $\varepsilon=0.5$.
For this mode the phase velocity $\text{Re} \omega(k)/k$ is extremely close to 1, but the precise
value has to be determined numerically. Since the damping rates of all
the three modes are very close, the above analytical result
nicely shows that the damping rate of the modes becomes arbitrarily
small for sufficiently large ${\sf Wi}=2{\sf S}$. Therefore, our amplitude
expansion becomes self-consistent for sufficiently large ${\sf Wi}$.

Once the  eigenmode has been obtained numerically, one has the
velocity and shear stress fields as a function of the coordinates. 
Fig.~\ref{velocityplot} of the introduction illustrates perturbed
velocity field in the planar case for ${\sf Wi}=8$ and $k=1.5$.

%\begin{figure}
%\begin{center}
%  \includegraphics[scale=.4,angle=-90]{phi0plplaatje.eps}
%\end{center}
%\caption[]{The eigenmode $\phi(\xi)$, ${\sf Wi}=8,k=1.2$.}\label{plaatje10}
%\end{figure}

Our nonlinear analysis in the next section will be based on using the
asymmetric mode which is the weakest damped. The reason for not basing
our expansion on the symmetric mode is twofold:\\
\hspace*{4mm} {\em (i)} The symmetry of the velocity $v_y$ of this
mode is such that it corresponds to a type of undulation mode between
the plates, which does not generalize easily to the case of a
cylindrical tube. \\
\hspace*{4mm} {\em (ii)}
The result  that the eigenvalue $\omega(k) = (1+ \delta)k -2i/{\sf
S}$ for this mode, with
$\delta$ a very small real quantity, implies that the factor $c(x)$
which appears at various places 
in the equations for the $\beta _i$ (see Appendix \ref{appendixa}), becomes very
large near the center. Consequently, the components of $\tau$ become very
large, almost singular,  near the center.
\\
\hspace*{4mm} {\em (iii)} The asymmetric mode {\em is} the one least damped.

\subsubsection{The cylindrical case}

\begin{figure}
\begin{center}
  \includegraphics[width=0.9\linewidth]{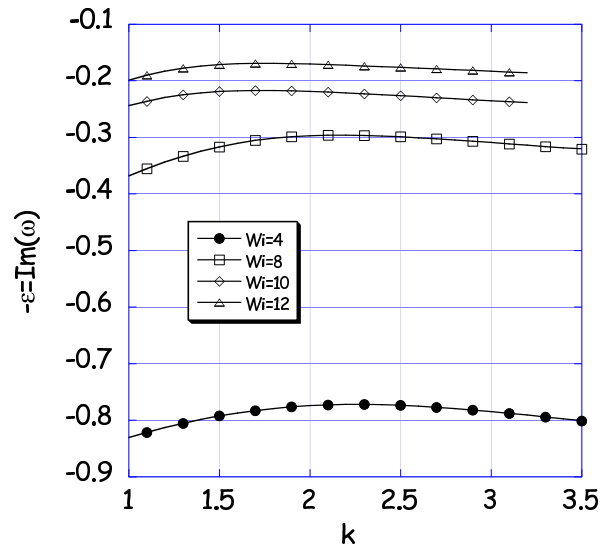}
\end{center}
\caption[]{Dispersion relation for the cylindrical case for ${\sf Wi}=4,8,
  10$ and $12$.}\label{plaatjecyl1}
\end{figure}

\begin{figure}
\begin{center}
  \includegraphics[width=0.9\linewidth]{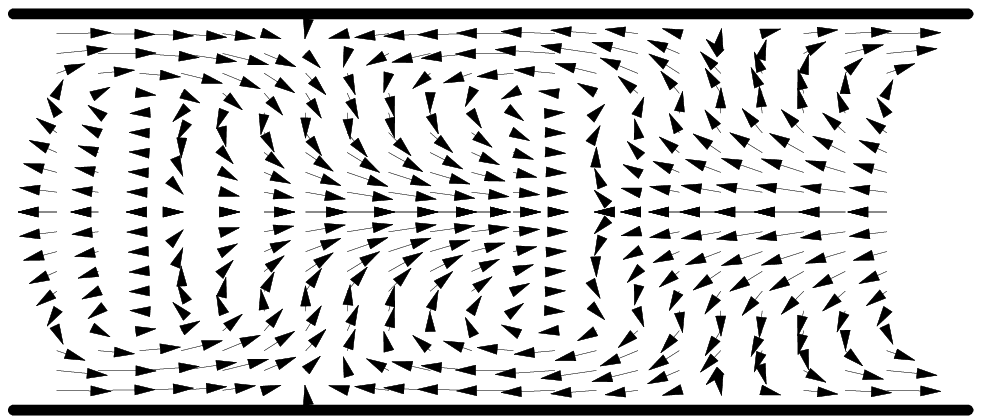}
\end{center}
\caption[]{Plot of the velocity field $\delta \vec{v}$ corresponding
to the linear eigenmode mode of the planar slit geometry with 
wavelength $\lambda=4 \pi/3$ at ${\sf Wi}=8$. The basic flow profile is in the
horizontal direction.}\label{velocityplot} 
\end{figure}

We now turn to  the numerical results for the cylindrical case.
Fig. \ref{plaatjecyl1} shows  the magnitude of the
imaginary part of the 
eigenvalue $\omega$   as a function
of $k$ and for various ${\sf Wi}$. 
As in the planar case,  all values of ${\sf Wi}$, $ \text{Im}\, \omega<0$. As the temporal behavior of the modes
is as $e^{-i\omega t}$, this again confirms 
that the flow is  {\em linearly stable}; the decrease of $\varepsilon$
with increasing ${\sf Wi}$ again confirms that the linear stability becomes
less and less with increasing flow velocity, roughly as $\varepsilon \sim 1/{\sf Wi}$.

\begin{figure}
\begin{center}
  \includegraphics[width=0.9\linewidth]{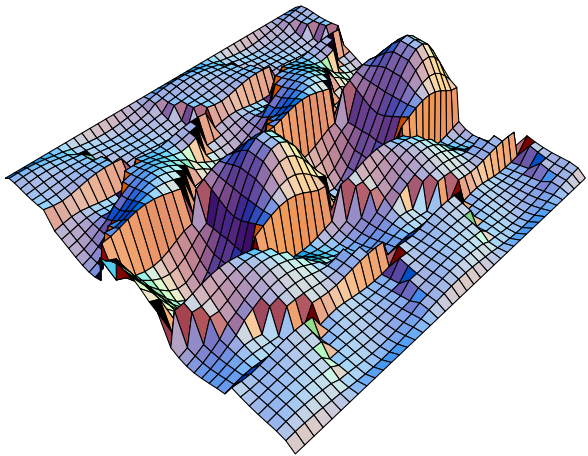}
\end{center}
\caption[]{The data of Fig.~\ref{velocityplotcyl} plotted
  differently. Here the component $\delta v_z$ is shown in a
  three-dimensional plot; it illustrates how the perturbation is
  largest at the center of the tube.}\label{velocityplotcyl2} 
\end{figure}

In Fig.~\ref{velocityplotcyl} we already showed the velocity profile
$\delta \vec{v}$ corresponding with the linear eigenmode with
wavelength  $\lambda=4\pi/3$ in the cylindrical
case. Fig.~\ref{velocityplotcyl} confirms that the same type of
behavior is found in the planar case; the main difference is that in
the cylindrical case, the mode is more confined close to the center of 
the tube (see also Fig.~\ref{velocityplotcyl2}) than in the planar
case. This is consistent with the fact
that, as we will see in Section IV, the nonlinearly most unstable mode 
has a shorter wavelength in the cylindrical tube than in the case of
the planar slit. In fact, the  roll-type
structure of the planar flow-profile is more evenly distributed over
the whole cross-section, its qualitative appearance is closer than the
cylindrical one to  the
flow pattern in the Taylor-Couette cell that  according to the arguments
of Groisman and Steinberg \cite{steinberg2} underlies   the
subcritical instability in that case.
\section{Nonlinear Analysis} \label{secna}

We will start with an outline of the method
that we use in the first section. As we shall see, in important
ingredient that we need to obtain to determine certain solvability
conditions is the solution of the linear adjoint operator.
We will discuss the derivation of the
adjoint problem in the sections 
\ref{secadj} and \ref{secsym}.
We finally discuss the numerical results in section \ref{secnum}.

\subsection{Outline of the structure of the amplitude expansion}
In order to perform the nonlinear analysis it is convenient to 
 rewrite the equations in a vector notation. We shall develop the
 framework for the planar case, and then indicate the changes in the
 cylindrical case at the end of this section. 

Since we only consider
 perturbations  which are independent of the coordinate $x$ in the
 planar case, the stress components $\tau_{xx}$, 
$ \tau_{xy}$, and $\tau_{xz}$ are all zero, also
in the nonlinear regime.
We can therefore capture the nonzero fields in  a five-component
vector $V$ whose  components  are $\delta v_y ,\delta v_z, \delta \tau
_{22} , \delta \tau _{23} $ and $ \delta \tau _{33}  $.
Furthermore, we rewrite equations (\ref{diveq}),(\ref{defpro}) and
(\ref{OBeq}) in the  
following form:
\begin{equation}
{\mathcal L} V=N(V,V), \label{fulleq}
\end{equation}
where ${\mathcal L}$ is the linear operator associated
with the linear problem of section \ref{secla}, and $N(V,V)$ contains
all the nonlinear terms. Thus the linear problem is simply ${\mathcal L} V=0$, and
$N(V,V)$ contains only terms from the left
hand side of equation (\ref{OBeq}); for the analysis below it is
convenient to take $N$ symmetrized, so that $N(V_1,V_2)=N(V_2,V_1)$.
Explicit expressions for $N(V,V)$ and ${\mathcal L}$ can be found in  Appendix \ref{appendixb}.

Normally, an amplitude expansion is performed about the critical point
where one of the modes is marginal (i.e., neither grows nor decays) 
for a particular value of the wavenumber $k$ \cite{ch,walgraef}. This
critical wavenumber 
corresponds to a maximum of the linear dispersion relation. In the
present case, the situation is not quite like this: the dispersion
relation does not have a clear maximum, and the linear modes are
always weakly damped. This difference is in practice not a great
problem. First of all, we will not be interested in spatial variations
of the envelope; instead, we just pick a particular mode $k$, and
determine the important one for our analysis at a later stage (through
the requirement that the threshold amplitude be minimal). It is
therefore not necessary to expand about a maximum of the dispersion
relation. Secondly, the amplitude expansion is based on an adiabatic
decoupling of the fast and slow scales. In this problem, the fast
scale is the period of the modes and the slow time scale is
associated with the linear decay time of the modes. We saw in section
\ref{secla}  that in dimensionless units, the frequency of the modes
is of order unity, while the damping rate $\varepsilon $ becomes much smaller than 1
for large enough ${\sf Wi}$. Thus, there is indeed a separation of times
scales, and this is essentially all that is needed for the analysis
below.

In fact, for readers experienced with amplitude expansions, we could
essentially go ahead pretending we are expanding about a true critical
point where the growth rate of one of the modes is zero, and then at
the end add the linear damping term by hand. Nevertheless, we prefer
to formulate the analysis  more carefully by keeping the damping term
as it stands.

A linear eigenmode of the equations is of the form
\begin{eqnarray}
V_0(\xi,z, t; k , \omega) & = & \tilde{V}_0 (\xi; k, \omega )\; e^{(ikz-i\omega t)} \nonumber \\
& = &  \tilde{V}_0 (\xi;
k, \omega)\; e^{(ikz-i \omega_rt)}\; e^{-\eps t}, \label{v0eq}
\end{eqnarray}
where we have introduced $\omega_r =\text{Re}\, \omega $. Of course, all components of
$V_0$ can just be obtained directly from the results of the linear
stability analysis 
of section \ref{secla}.

In an amplitude equation formalism, one can in general also allow for
spatial variations of the amplitude on a slow scale. As we already
remarked above, here we confine the analysis to the temporal evolution
of a single mode $k$ and its harmonics. We do so for two reasons:
First of all, it 
simplifies the analysis, secondly, our main goal is to determine
whether there is a weakly nonlinear instability. Anticipating that we
will find that there is one, there is then not much to be gained 
in allowing for slow spatial variations.

Our aim thus is to study the weakly nonlinear evolution in time of the
amplitude $A$ of a
single mode of the form
\begin{equation}
 A(t) \; \tilde{V}_0 (\xi;
k, \omega)\; e^{(ikz-i\omega_rt)} + c.c.  \label{amode}
\end{equation}
Of course, such a single mode is not really a solution of the (weak)
nonlinear equation, but the corrections are automatically accounted
for in the amplitude equation formalism and are slaved to (\ref{amode}).
We also note that because of the normalization (\ref{bc2}),
$\phi''(1)=1$, we have 
\begin{equation}
\mbox{max}\left[\partial_\xi \delta v_z(\xi)|_{\text{wall}}\right] = 2 |A| 
\end{equation}
which, because in dimensionless units the shear rate at the wall in
the unperturbed flow field equals 2, leads immediately to the  the
relation (\ref{amplitude}) between the maximum relative shear
rate distortion at the wall and the amplitude $A$. Likewise the
relation (\ref{stressamp}) between the amplitude $A$ and the maximum
shear stress distortion at the wall follows from the explicit
linearized dynamical equations. These identifications are  
important in translating our results to real values, as was
already noted  in section I.C and Fig.~\ref{plaatje4} and \ref{plaatje5}.

We want to know in particular whether the amplitude $A$ will grow or
decay in time 
when the nonlinearities are taken into account. In the amplitude
equation formalism, $A(t)$ is expected to  depend only on the slow timescale $\eps
t$, but we have made this explicit, for  reasons that become clear
below. We have 
to keep track of the fact that the true zero mode of our linear
operator ${\mathcal L}$ is $V_0$ as given in (\ref{v0eq}). This term includes the
weakly damped exponential factor $e^{-\eps t}$, and this term is not
explicit in (\ref{amode}). The amplitude equation now proceeds by
constructing the weakly nonlinear solution by writing \cite{ch,walgraef}
 \begin{equation} \label{sub8}
V =  \eps^{\frac{1}{2}} {V}^s_0  + \eps V_1 +  \eps^{\frac{3}{2}} V_2
+ \cdots, 
\end{equation}
where $V^s_0$ is the real quantity defined as the sum
\begin{equation}
{V}^s_0 = B(T) V_0 + B^{\star}(T) V_0^{\star},
\end{equation}
and where in writing $B(T)$ we have anticipated that the amplitude $B$
varies on the \revision{slow timescale}
$T = \eps t$.
Note that the real combination ${V}^s_0$  is needed because the
vector $V$ refers to real physical fields.
Once we have derived the equation for $B$, comparison with
(\ref{amode}) shows that we should
make  the association
\begin{equation}
A(t)  \Longleftrightarrow \eps^{\frac{1}{2}} B(T) e^{-\eps t}. \label{AtoB}
\end{equation}
The other terms in (\ref{sub8}) are then precisely the corrections to
the dominant mode which are necessary to construct a weakly nonlinear
solution.

In the subsequent analysis, we have to keep in mind that $V_0$, with the temporal factor included,  is the
true zero eigenmode of the linear operator ${\mathcal L} $: The temporal
derivatives in the linear operator in fact work explicitly on both
temporal exponential factors in the expression (\ref{v0eq}) for
$V_0$. When substituting the form (\ref{sub8}) in the equations, we
then also have to account for the time derivatives of $B(T)$. For a
product term of the form $B(T) e^{-i\omega t}$ we then have
\begin{equation}
\frac{\partial \left( B(T) e^{-i\omega t} \right)}{\partial t} = B(T)
\frac{\partial e^{-i\omega t}}{\partial t} + e^{-i\omega t} \eps \frac{\partial B(T)}{\partial T},
\label{tderivs}
\end{equation}
which we can simply summarize by making the substitution
$\de _t \rightarrow \de _t + \eps \de _T$
in all the time derivatives in the linear operator.
\revision{Since only first order time derivatives enter the linear operator ${\mathcal L}$,
we get an expansion of ${\mathcal L}$ of the form}
${\mathcal L} = {\mathcal L}_0 + \eps {\mathcal L}_{T}$.

The amplitude expansion now proceeds by substituting the expansions
 for ${\mathcal L}$ and (\ref{sub8}) for $V$ in the nonlinear
equation (\ref{fulleq}) and collecting the terms order by order in $\eps$.
For the first three orders in $\eps$ we get:
\begin{eqnarray} \label{sub12}
\mbox{order}~\eps^{\frac{1}{2}}:\hspace*{25mm} {\mathcal L}_0 {V}^s_0& = & 0,
\\
\label{sub13}
\mbox{order}~\eps^{1}:\hspace*{25mm} {\mathcal L}_0 V_1 & =&  N({V}^s_0,{V}^s_0),
\\
 \label{sub14}
\mbox{order}~\eps^{\frac{3}{2}}:~~~~~~~~~~{\mathcal L}_0 V_2 +  {\mathcal L}_{T} {V}^s_0 & = &
N({V}^s_0,V_1). 
\end{eqnarray}
Equation (\ref{sub12}) is satisfied identically, as it should, because
${V}^s_0$ is the sum of two terms with wavenumber $k$ and $-k$ which
are both zero eigenmodes of 
${\mathcal L}$. The components of the $V_0(\xi)$
follow from the numerical results for 
$\phi$  of section \ref{secla}.
We can now solve equation (\ref{sub14}) to find $V_1$.
This  gives us
\begin{equation}
V_1 =  B^2(T,t) \, e^{2i(kz+\omega_r t)-2\eps t} \tilde{V}_1(\xi) + C(T) \, V_0
~+~c.c.
\end{equation}
where we have to find the vector $\tilde{V}_1(y)$ numerically,
because N(V) contains only quadratic terms. The second term  is
allowed with arbitrary $C$, since $V_0$ is a zero mode 
of the linear operator ${\mathcal L}_0$; it will not be needed in the subsequent
analysis. In principle, there would also have to be an additional
$k=0$ term, but this term is identically zero for the case of 
the asymmetric boundary conditions of interest here.
Once $V_1$ is known, we can proceed to Eq. (\ref{sub14}); in this
equation, the operator ${\mathcal L}_T$ works on $B(T)$ and its complex conjugate
only: we can write the equation as
\begin{equation} \label{sub16}
{\mathcal L}_0 V_2 + ( V_0 \; {\mathcal L}_T B + V^*_0\; {\mathcal L}_T B^*) = N(V_0,V_1).
\end{equation}
This equation determines the time derivative of $B$ through a
solvability condition:  since the operator ${\mathcal L}_0$ has a right zero
mode, it can be solved if and only if the other two terms in the
equation are orthogonal to the left zero mode of ${\mathcal L}_0$
\cite{ch,walgraef}. This
requirement gives us the desired equation of ${\rm d}B/{\rm d}T$. To make this
explicit, we first have to define the adjoint problem and an inner product.

%\subsubsection{The solvability condition}

We define a space of 5-dimensional vector functions of the three
variables $\xi, z$ and $t$,
\begin{equation}
\Omega = \{ f: R^3 \rightarrow R^5 :f=\tilde f (\xi,t) e^{i(kz-\omega_rt)} \}.
\end{equation}
The components of the functions $f$ satisfy the physical boundary
conditions discussed in section \ref{secla}. Furthermore, an 
 inner product $I$ on $\Omega$ is defined:
\begin{eqnarray}
I(w,f)=\frac{1}{2} \int _{-1}^{1}\!\!\!{\rm d}\xi\; \frac{2\pi}{k} \int_0
^{\frac{2\pi}{k}}\!\!\!{\rm d}z \; \frac{2 \pi}{\omega_r} \int _0 ^{\frac{2
\pi}{\omega_r}}\!\!\!{\rm d}t \otimes \nonumber \\ \otimes \sum_{i=1}^5 w^{\star}_i(\xi,z,t)f_i(\xi,z,t).\label{Iexpression}
\end{eqnarray}
On this space of functions, an adjoint operator ${\mathcal L}_0^\dagger$ is
defined such that 
\begin{equation} \label{sub33}
I(w,{\mathcal L}_0f) = I({\mathcal L}_0^{\dagger}w,f)
\end{equation}
for every function $f$ in $\Omega$. Because the adjoint operator is
obtained through partial integrations with respect to $\xi$, the
requirement that (\ref{sub33}) does not yield any boundary terms from
these partial integrations yields the appropriate boundary conditions for the
functions $g$ in the adjoint space. We will state these explicitly
for our case in section \ref{secadj}. 

Let $W_0$ be the zero mode of ${\mathcal L}_0^\dagger$.  The solvability
condition applied to (\ref{sub16}) then becomes
\begin{equation} \label{sub30}
I(W_0,N(V^s_0,V_1)- V_0{\mathcal L}_T B -V^{\star}_0 B^{\star} ) = 0.
\end{equation}
Let us focus on the term with $B$; if we write 
\begin{equation}
W_0 = \sum _{m \in \mathbb Z} g_m(y,T,t) e^{im (kz -\omega_r t)}
\end{equation}
we see that we only need the term with $m=1$, since the inner product
in (\ref{sub30}) is seen to vanish
for all other terms after the $z$-integration is performed.
For the same reason, we  only need the terms which are proportional to
$\expor$ from  $N(V_0,V_1)$.
Using this we obtain the following equation for the derivative of $B$:
\begin{equation} \label{sub32}
\frac{1}{2} \int _{-1} ^1\!\!\!{\rm d}\xi\;  \left(W_0 ^{\star}, -{\mathcal L}_T B + |B|^2 B N(V_0 ^{\star},V_1)\right) = 0.
\end{equation}
The complex conjugate of this equation is obtained from analyzing the
term with $B^*$ in (\ref{sub30}).

In (\ref{sub30}),  we have used the approximation that $\eps$ is
sufficiently small that we are allowed to write
\begin{equation}
\int_0^{\frac{2\pi}{\omega_r}}\!\!\!{\rm d}t\; B(T) e^{-\eps t}\, (\cdot)  = B(T)
e^{-\eps t}  \int_0^{\frac{2\pi}{\omega_r}}\!\!\!{\rm d}t\; (\cdot). 
\end{equation} 
This is nothing but the usual adiabaticity assumption. As we have
shown in section \ref{secla}, this approximation is justified for
large ${\sf Wi}$. 
 We can summarize  the Eq. (\ref{sub32})  in the form
\begin{equation}
\de _T B = c_0 |B|^2 B \label{Bfinal},
\end{equation}
where $c_3$ is a complex quantity which is obtained from working out
the two inner product terms.
We can now translate this result back into the lowest order nonlinear
equation for the amplitude $A$. Upon combining (\ref{AtoB}),
 and (\ref{Bfinal}), we finally get
\begin{equation}
\de _t A = - \eps A + c_3 |A|^2 A,
\end{equation}
which is nothing but Eq. (\ref{Atriple}) of the introduction in dimensionless
units. As discussed there, the sign of the real part of $c_3$
determines whether or not we are dealing with a subcritical bifurcation.

In the following sections the boundary conditions of the adjoint
problem are derived, and we then proceed to solve adjoint problem and
to determine $c_3$.

The framework laid out above can be extended rather easily to the case 
of the cylindrical tube. In that case we consider axially symmetric
perturbations only, so that both  $\tau_{\theta r}$ and 
$\tau_{\theta z}$ vanish; note, however, that 
$\tau_{\theta \theta}$ is nonvanishing in this case.  The
vector $V$ therefore now has six components, $\delta v_r , \delta v_z , \delta \tau
_{rr} , \delta \tau _{rz} , \delta \tau _{\theta \theta} $ and $ \delta \tau_{zz} $.
Apart from this change, the structure of the expansion is essentially
the same, except for trivial changes like the fact that the first 
integration in (\ref{Iexpression}) should now be taken over the two-dimensional
scaled radial coordinate $\xi$.

\subsection{The adjoint operator and associated boundary conditions}
\label{secadj} 

In this section we will calculate the components of the
adjoint operator ${\mathcal L}_0^\dagger$ using the defining equation
(\ref{sub33}). We will follow again the planar case, and indicate the
major changes for the cylindrical case at the end an in the
appendices.
Writing $V=(v_1,\ldots,v_5)$ and $W=(w_1,\ldots,w_5)$ we have
\begin{equation} {\mathcal L}_0V =  \left( \begin{array}{ccccc}
 \de_y v_1+\de_z v_2 \\
 \de _y \de _z v_3 + 
  (\de _z ^2 - \de _y ^2)v_4 -\de _y \de _z v_5 \\ Cv_1+Av_3 \\ Dv_1+Ev_2+Bv_3+Av_4 \\ Fv_1+Gv_2+2Bv_4+Av_5 \end{array} \right).
\end{equation}
The various functions and coefficients in this expression are given in
Appendix \ref{appendixb}.
We will illustrate the structure of  the calculation by analyzing two
terms of $I(W,{\mathcal L}_0V)$, and simply state the 
results for the other terms.
One term we get is $w_3 ^{\star} A v_3$:
\begin{eqnarray} \label{amp8}
\int \!\! w_3 ^{\star} A  v_3 
&=&  \int \!\! w_3 ^{\star} \left(1+\de _t +\frac{{\sf S}}{2} v_z^0(\xi) \de _z\right) v_3 , \nonumber \\
 &=& \int \!\!\left(1-\de _t - \frac{{\sf S}}{2} v_z^0(\xi) \de _z\right) w_3 ^{\star} v_3  \nonumber \\
 & & ~~~+ \frac{{\sf S}}{2} \int \!\!\de _z  (w_3 ^{\star} v_3) + \int \!\!\de _t (w_3 ^{\star} v_3).
\end{eqnarray}
Both integrals on the last line vanish, the first one because the 
integrand is a partial $z$-derivative of a term which is periodic in $z$, the
second one because the integrand is a partial $t$-derivative of a term
which is periodic in $t$. Thus we simply obtain
\begin{equation}
\int \!\!\!w_3 ^{\star} A v_3 = \int \!\!\! \left(1-\de _t -
\frac{{\sf S}}{2} \vzy \de _z\right) w_3
^{\star} v_3. 
\end{equation}
In short: we pick up a minus sign for every partial integration and in
performing the partial integrations we get, in general, boundary terms
which have to vanish. In the above example, the boundary terms
trivially vanished because of the periodicity of the terms with
respect to $t$ and $z$, but the partial integrations with respect to
$\xi$ do not automatically vanish. In particular, from the various
terms we get the following boundary conditions:
\begin{eqnarray} \label{amp5}
w_3 ^{\star}v_1(1)-w_3 ^{\star}v_1(-1)&=&0, \nonumber \\
w_4 ^{\star}v_1(1)+w_4 ^{\star}v_1(-1)&=&0, \nonumber \\
w_4 ^{\star}v_2(1)-w_4 ^{\star}v_2(-1)&=& 0, \nonumber \\
 w_5 ^{\star}v_2(1)+w_5 ^{\star}v_2(-1)&=&0, \nonumber \\ w_1
^{\star}v_1(1)-w_1 ^{\star}v_1(-1)&=&  0.
\end{eqnarray}
and
\begin{eqnarray} \label{amp6}
w_2 ^{\star} \de _z v_3(1)-w_2 ^{\star} \de _z v_3(-1)&=&0, \nonumber \\
w_2 ^{\star} \de _y v_4(1)-w_2 ^{\star} \de _y v_4(-1)&=&0, \nonumber \\ 
\de _y w_2 ^{\star} v_4(1)-\de _y w_2 ^{\star} v_4(-1)&=&0, \nonumber \\
w_2 ^{\star} \de _z v_5(1)-w_2 ^{\star} \de _z v_5(-1)&=&0.
\end{eqnarray}
The first set of conditions  (\ref{amp5}) is always
satisfied, because 
$V$ satisfies the boundary conditions of the original problem: $v_1(\pm 1)=v_y(\pm 1)=0$ and $v_2(\pm 1)=v_z(\pm 1)=0$.
By setting  $w_2 =0$ at $\xi =\pm 1$,
we have only one condition left of the second set:
\begin{equation}
\de_\xi (1) w_2 ^{\star} v_4(1)-\de _\xi w_2 (-1) ^{\star} v_4(-1)=0.
\end{equation}
In order to understand how many boundary conditions we need to impose,
it is good to realize that
the operator $\mathcal L$ actually works on physical, and hence real functions;
thus, we need the real
part of this combination to be zero. Thus,
we can still choose the phases of both amplitudes:
we use this freedom to set the phase difference between
the adjoint solution $W_0$ and the solution $V$ such that 
the combination $w_2 ^{\star} v_4(1)$ becomes purely imaginary.
This shows us that if we impose
\begin{equation}
\text{Re} \left(\de_\xi w_2 ^{\star}|_1 v_4(1)-\de _\xi w_2 ^{\star}|_{-1} v_4(-1)\right)=0
\end{equation}
 all conditions are satisfied for physical functions.
We can give all $w_i$ in terms of $w_2$ and its derivatives.
This gives a fourth order equation for $w_2$ which we
solve using a shooting method. We will discuss
the appropriate symmetry conditions in the next section.

In the cylindrical case, we need $w_2$ to be bounded. In the present case, this  can be 
achieved  \revision{by putting}
$ w_2(0)=0$, $w_2''(0)=0$.
These conditions  are satisfied if we expand
$ w_2(\xi)=a_1 \xi + \frac{1}{6} a_3 \xi ^3$, for $0<x <0.01$.
Furthermore we can set $a_1 = 1$ to eliminate the degree of freedom
we have. This leaves us with one shooting parameter, $a_3$.
We again have \revision{the boundary condition that $w_2$ vanishes on the wall},
$w_2(1)=0$.
We choose $a_3$ such that this condition is satisfied.

\subsection{Symmetry of the solution and the adjoint solution} \label{secsym}
As we discussed in section section \ref{secla}, the linear planar mode that we investigate is
asymmetric, which means that its symmetry is
\begin{equation}
\tilde{V}(\xi) = (odd,even,even,odd,even) \label{v0symmetry}
\end{equation}
which can be verified, see Appendix \ref{appendixa}.
For the adjoint solution we have also have two zero modes with
different symmetry,
\begin{equation}
\tilde{W}^{(1)}_0(\xi) =  (even,odd,even,odd,even) ,
\end{equation}
and 
\begin{equation}
\tilde{W}^{(2)}_0(\xi) =
 (odd,even,odd,even,odd).
\end{equation}
It turns out that for our choice of the linear mode
(\ref{v0symmetry}), the second mode $W^{(2)}_0$ does give a
solvability condition. This can be seen as follows. Only the last
three components of the vector $V$ contain a derivative $\de _T$. Thus
only the overlap of the third, fourth and fifth component of $W_0$ and
$V_0$ come in; because of the symmetry of the opposite symmetry of the
components, all these integral vanish. The same holds for the
$N$-term, and hence the solvability condition is trivially satisfied. 
As a result, $W^{(1)}_0$ is the adjoint zero mode which gives the
nontrivial boundary condition. The oddness of the component $w_2$ is
then guaranteed by taking
\begin{equation}
w_2(0)=0, \quad w_2'(0)=1,\quad w_2''(0)=0. \label{askbernard}
\end{equation}
These conditions, together with the boundary condition
at $\xi=\pm 1$ completely fixes the adjoint zero
mode. As with the linear problem discussed in section \ref{secla}, we
solve the differential equation together with the boundary conditions
with a shooting method.

In the cylindrical case, we have just one single mode, also for the
adjoint problem; the boundary conditions   that we
already derived above for the cylindrical case are completely
analogous to the above boundary conditions (\ref{askbernard}).

 \subsection{Numerical Results} \label{secnum}
In summary, the nonlinear term $c_3$ whose real part governs the weak
nonlinear stability is obtained numerically as follows.
We first solve the fourth order equation for the stream function
$\phi_0$, derived in section \ref{secla}.  This gives us the
components of the vector $V_0(\xi)$.
A second  routine of this program generates the term $N(V_0,V_0)$ too.
A second program  calculates the stream function $\phi _1$ of the
inhomogeneous equation (\ref{sub13}) from which we
obtain the components of the vector $V_1(y)$.
A third program solves the adjoint problem and gives the
vector $W_0$. 
Because we have $W_0 \sim e^{i(kz-\omega_rt)}, V_0 \sim e^{i(kz-\omega_rt)}$ and
$V_1 \sim e^{2i(kz+\omega_r t)}$ we have trivial $z$ and $t$
integrations in the inner product (\ref{sub32}).
The $\xi$-integration is  done numerically
from $0$ to $1$ because of symmetry. This then gives us the
coefficient $c_3$.

\begin{figure}
\begin{center}
  \includegraphics[width=0.9\linewidth]{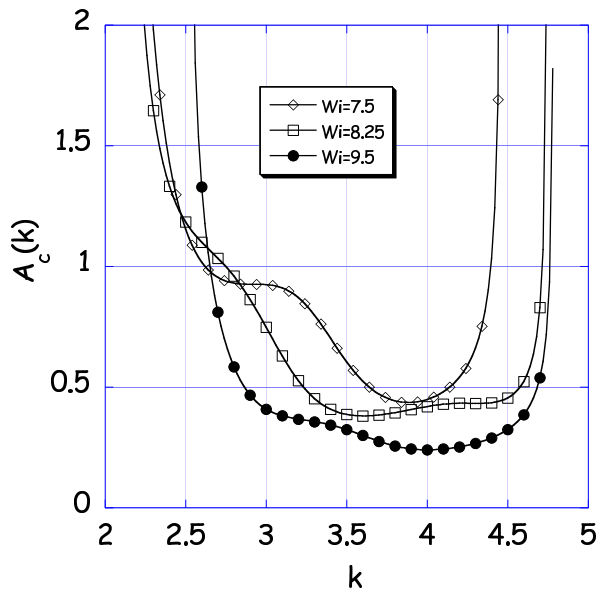}
\end{center}
\caption[]{The value of the critical value $A_{\text{c}}$ as a function of the
wavenumber $k$  for three different values of the Weissenberg number
${\sf Wi}$ for the case of the cylinder.}\label{f10} 
\end{figure}

We first discuss the cylindrical case. In Fig.~\ref{f10}  we plot the
value of the critical amplitude $A_{\text{c}}(k)$, determined from $c_3$ by
Eq.~(\ref{criticalamp}), as a function of the wavenumber  $k$
for three different values of the Weissenberg
number. The curves illustrate that for ${\sf Wi} > {\sf Wi}_{\text{c}}$  there is
a band of wavenumbers where $\text{Re}\, c_3 >0$ and that hence  there is, in
our approximation, a critical amplitude beyond which the flow is
unstable. The sharp rise of the critical amplitude at the edges of the
bands shown in Fig.~\ref{f10}  are close to the edges of the band
where $\text{Re}\, c_3 \to 0$.  Note also that for ${\sf Wi} =7.5$, the critical
value still has a rather sharp minimum near $k=4$, but that for
increasing values of ${\sf Wi}$ the bottom of the band flattens rapidly. For
decreasing values of ${\sf Wi}$, especially when ${\sf Wi}$  approaches about 5,
the band likewise sharpens; we will see this from a different
perspective below. 

\begin{figure}
\begin{center}
  \includegraphics[width=0.9\linewidth]{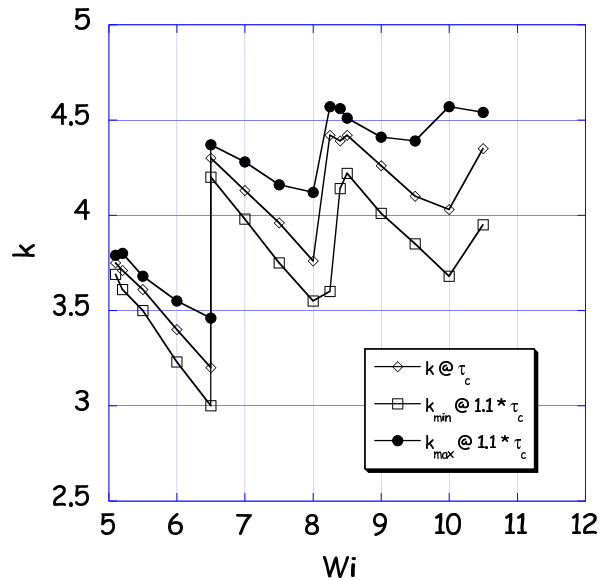}
\end{center}
\caption[]{The value of $k$ corresponding to the minimum value of the
curves in Fig.~\ref{f10} as a function of ${\sf Wi}$, and the values of $k$
at which the amplitude $A$ is 1.1 times the minimum value. The jumps
of the curves are due to fact that a curve of $A_{\text{c}}$ as a function of
$k$ has several minima. }\label{f11} 
\end{figure}

Fig.~\ref{f10} also shows the behavior of the critical amplitude
$A_{\text{c}}(k)$ as a function of $k$ shows rather complicated structure. For
${\sf Wi}=7.5$ there is a plateau in the critical value of $A_{\text{c}}(k)$ around
$k=2.7$; upon increasing ${\sf Wi}$, this plateau shifts and disappears, 
 whereas the absolute minimum of the curve shifts to smaller 
$k$ values while a new minimum develops at larger $k$. Already at
${\sf Wi}=8.25$, 
the two minimum almost correspond to the same values of $A_{\text{c}}$,  but 
for slightly larger ${\sf Wi}$ the minimum at larger $k$ value 
becomes the absolute minimum. This is further illustrated in
Fig.~\ref{f11}, where we plot the values of $k$ corresponding to the
absolute minima of the curves, as well as the values corresponding to
an amplitude 1.1 times the minimum value, as a function of ${\sf Wi}$. The
figure illustrates that between ${\sf Wi}\approx 6.5$ and ${\sf Wi} \approx 8$,
the minimum of the curve shifts to smaller $k$ values, but that around
${\sf Wi}\approx 8$ a local minimum at higher $k$-values becomes the
absolute minimum. 

\begin{figure}
\begin{center}
  \includegraphics[width=0.9\linewidth]{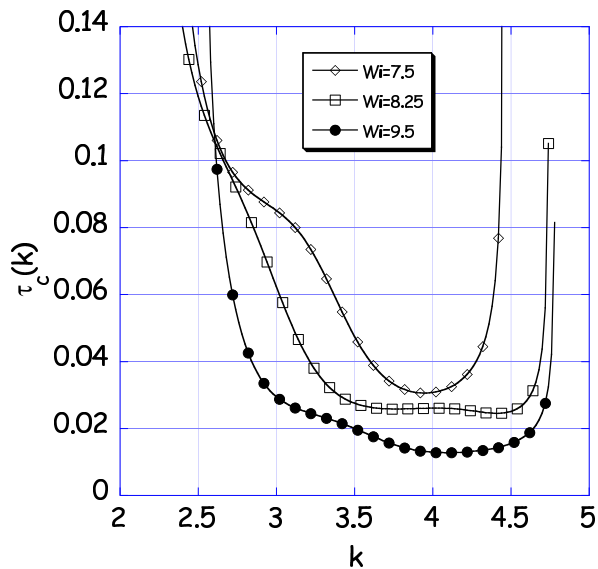}
\end{center}
\caption[]{The values of the dimensionless shear stress $\tau$ beyond
which the flow is unstable in the cylinder, as a function of $k$ for
the same three values of ${\sf Wi}$ as in Fig.~\ref{f10}. }\label{f12} 
\end{figure}

As explained in section \ref{summary}, our normalization is such that $A$ is the
maximum perturbation in the shear rate at the wall over one
wavelength, divided by the shear rate at the wall of the unperturbed
parabolic profile [See Eq.~(\ref{amplitude})].  Upon increasing ${\sf Wi}$,
the minimum values of the curves, which were already plotted in
Fig.~\ref{plaatje5}, quickly decrease.  As we already mentioned in the
introduction, we have also analyzed the critical value for the
relative shear stress perturbation at the wall beyond which the flow
is unstable [see Eq.~(\ref{stressamp})]. The data for this ratio as a
function of $k$ for the same values of the Weissenberg number as in
Fig.~\ref{f10} are shown in Fig.~\ref{f12}.  There are several  important
things to note about this figure. First of all, the curves of the
critical shear stress perturbation have just one minimum, contrary to
those for the critical shear perturbation, and secondly, the values
for the critical shear stress perturbation are typically a factor ten
smaller than those for the critical shear. As Fig.~\ref{f12} shows,
for ${\sf Wi}=9.5$ a perturbation of about 1\% in the wall shear stress is
sufficient to render the flow unstable. This is why in
Fig.~\ref{plaatje5} the values of the critical shear stress amplitude
(the values corresponding to the minima of the curves in
Fig.~\ref{f12}) are multiplied by 10 to draw them on the same scale as
$A_{\text{c}}$.  Finally, we note that the edge of the band of unstable shear
stress perturbations $\tau$ is the same as the edge of the band of
unstable shear perturbations --- this is simply due to the fact that
the edge of the band is marked by the point where $\text{Re}\, c_3=0$.

\begin{figure}
\begin{center}
\includegraphics[width=0.9\linewidth]{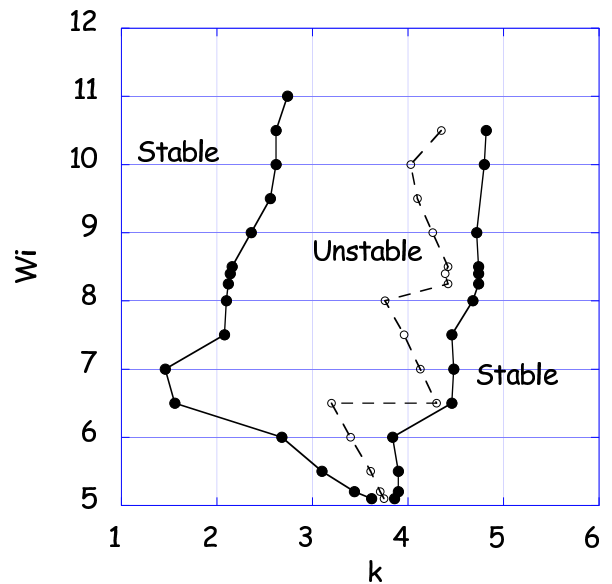}
\end{center}
\caption[]{Plot of the width of the $k$-band where the corresponding
modes render the basic cylindrical Poiseuille flow nonlinearly
unstable. The dashed line indicates the $k$-value of the mode which is
most unstable to shear perturbations, i.e. the minimum of the curves
plotted in Fig.~\ref{f10}.}\label{f13} 
\end{figure}

In Fig.~\ref{f13} we plot the band in which modes are nonlinearly
unstable beyond some critical value as a function of ${\sf Wi}$. This figure
clearly illustrates that the width of the band vanishes at ${\sf Wi} =
{\sf Wi}_{\text{c}}\approx 5$, since below this values $\text{Re}\, c_3 < 0$ for all
$k$.  This figure is somewhat reminiscent of  the so-called  ``Busse
balloon'' of  Rayleigh-B\'enard convection \cite{ch}, but one should
keep in mind that the interpretation is slightly different as we are
dealing with a subcritical (inverted) bifurcation. Thus, the regions
marked ``Stable'' indicate the regions in the diagram where the
perturbations with a wavenumber $k$ in that region are nonlinearly
stable in our approximation. The basic Poisseuille flow profile is,
however, nonlinearly unstable at all ${\sf Wi}_{\text{c}}$ to modes whose
wavenumber lies in the band marked ``Unstable''. In reality, we expect
that the flow pattern will settle to some kind of stable nonlinear
behavior dominated by modes in this band.

Although the band of unstable modes widens toward small $k$-values for
${\sf Wi}$ between 6 and 7.5 --- a feature which is related to the plateau
in Fig.~\ref{f10} for ${\sf Wi}=7.5$ --- at flow rates (Weissenberg numbers)
about 50\% beyond the critical value, the band of unstable wavenumbers
ranges from just over 2 to a value somewhat larger than 4. Although
with our expansion to cubic order in the amplitude we can not probe
the stable nonlinear patterned flow regime nor the nonlinear selection
of the wavenumber (it need {\em not} correspond to the most unstable
mode), it seems reasonable to 
assume that the wavenumbers of the pattern in the tube will lie in the
range identified by Fig.~\ref{f13}. 

Because near ${\sf Wi}_{\text{c}}$ the critical amplitude is
relatively large, and again because we have only expanded up to cubic
order, one has to interpret our results near ${\sf Wi}_{\text{c}}$ with
caution. Nevertheless, our results appear to identify the location of  the  saddle-node
bifurcation to nontrivial flow patterns near ${\sf Wi}_{\text{c}}\approx
5$. Since the band of unstable modes in this limit is very small, it
appears likely that if one follows the nontrivial flow pattern down to
this value, its wavenumber is should be close to the value where the
band in Fig.~\ref{f13} closes,
\begin{equation}
{\sf Wi} \approx {\sf Wi}_{\text{c}} \approx 5: ~~\Longrightarrow~~~~k\approx 3.75,
\end{equation}
since the basic profile is then nonlinearly stable to perturbations
with wavenumber that differs significantly from this value.
In dimensional units, this implies for the wavelength $\Lambda$ of the
pattern
\begin{equation}
{\sf Wi} \approx {\sf Wi}_{\text{c}} \approx 5: ~~~~~~\Lambda\approx
2 \pi R/ 3.75 \approx  1.7 \, R.
\end{equation}

It is important to keep in mind that $k$ is the wavenumber {\em inside} the
cylindrical tube. If the (near)periodic flow and stress patterns
inside the tube indeed cause  the  surface undulations  that are the
first stage leading to melt fracture outside the
tube at larger flow rates, then one
should keep in mind that $k$ is {\em not} the wavenumber of these
surface undulations. For, the polymer extrudate swells upon leaving
the tube, and the flow velocity profile of the polymer  becomes
essentially constant after leaving the tube. A better way to compare
our theory with measurements on the extrudate is therefore to compare
the (dominant) frequency of the undulations: since no oscillations
will disappear at the outlet  of the tube, the frequency measured inside
the tube must be equal to the frequency with which  the 
the extrudate width oscillates after flowing out of the tube \cite{bonn1}. 
Of course, the frequency of the oscillations on the nonlinear flow
branch corresponding to the solid line in Fig.~\ref{figconjecture} can not be
obtained precisely from our expansion; assuming that the
nonlinearities do not change this frequency too much, we estimate it
from our results for the linear modes. From our analysis of the linear
eigenmodes in section \ref{secla} we find that the dimensionless
frequency $\text{Im}\, \omega$ of the modes is  to a very good
approximation (better than to a percent or so) given by
\begin{equation}
\text{Im}\, \omega = k -0.93/{\sf Wi}.
\end{equation}
This result implies that for large Weissenberg numbers, the phase
velocity $\text{Im}\, \omega /k$ approaches 1. 
Since according to (\ref{dimensionlesscyl}) we measure velocities 
 in units of $v_{\text{max}}$, this means that 
the periodic stress pattern moves essentially with the maximum
velocity of the flow in the large Weissenberg limit. This indicates
that the linear mode  more and more concentrated near the  axis
of the tube for large ${\sf Wi}$. 
The above result gives as an estimate for the frequency $f$ in
dimensional units 
\begin{equation}
f = v_{\text{max}} (k-0.93 /{\sf Wi}) /(2 \pi  R )
\end{equation}
Finally, using that we expect  the transition to occur at ${\sf Wi}_{\text{c}}\approx
5$ and $k\approx 3.75$, and that for the unperturbed (parabolic)
profile of a UCM fluid $v_{\text{max}}= 2 v_{\text{av}}$ with
$v_{\text{av}}$ the average flow velocity in the tube, our estimate
for the frequency at the transition becomes
\begin{equation}
\mbox{at transition:}~~~ f \approx  1.13 \, v_{\text{av}} /R . \label{transfreq}
\end{equation}
It is important to realize that this estimate is based on the
assumption that the frequency  (or phase velocity) does not get
renormalized significantly by the nonlinearities. This is maybe not
unreasonable near threshold, where the amplitude of the periodic
modulation of the pattern will be smallest (keep in mind that since we
are dealing with a subcritical bifurcation, the amplitude near
threshold remains finite). 

\begin{figure}
\begin{center}
 \includegraphics[width=0.9\linewidth]{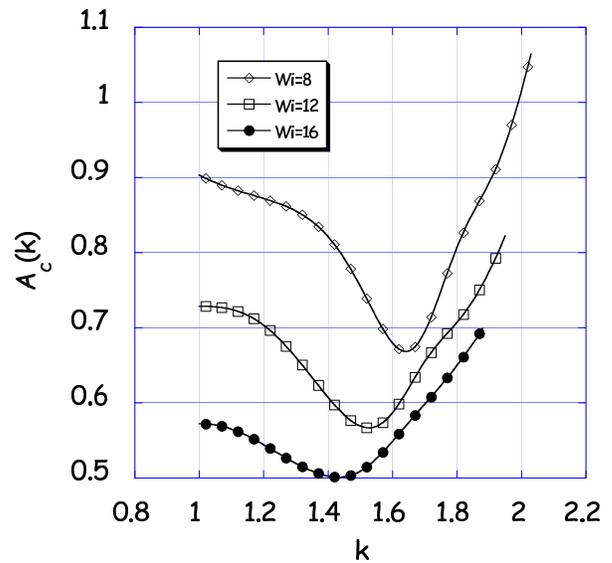}
\end{center}
\caption[]{Plot of the critical amplitude $A_{\text{c}}$ as a function of $k$
in the planar case for three different values of ${\sf Wi}$.}\label{f14}
\end{figure}

Especially further above threshold,  we
expect the renormalization of the frequency to be
significant. Although this can not be calculated from our first
nontrivial nonlinear term in the amplitude expansion, we can
reasonably estimate the frequency at values of ${\sf Wi} $ of the order of
50 to 100\% above threshold, say, as follows.  As we noted above, in
this range, the band of unstable wavenumbers $k$ ranges from just over 2
to about 4.5 (See Fig.~\ref{f14}). It is unlikely that the range of wavenumbers will change
drastically due to nonlinear interactions, since in according to our
calculations outside this range both the linear and the cubic term in
the amplitude expansion are stabilizing. On the other hand, the
frequency of oscillations, if one studies the pattern in a fixed lab
frame, will get strongly renormalized: once the pattern gets well
developed, we  expect (but have no proof) that the nonlinearly
modulated profile moves with the {\em average speed} $v_{\text{av}}$
rather than with the maximum speed of the unperturbed parabolic
profile (keep in mind that in linear order, the periodic linear eigenmode does
not affect the average speed, as it averages out to zero over one
wavelength, but that this does not remain true in nonlinear
order). Which particular wavenumber will be selected nonlinearly (or
whether in fact a well-defined sharp wavenumber will be selected nonlinearly
in a carefully controlled experiment) we do not
know, but with the assumption that it lies in the unstable band $
k_{\text{min}} < k < k_{\text{max}}$ and
that it moves with the average speed, we get the following estimate
for the dimensional frequency\cite{notefreq}
\begin{equation}
\frac{k_{\text{min}} v_{\text{av}}} {2\pi R}  \lesssim f \lesssim
\frac{k_{\text{min}} v_{\text{av}}} {2\pi R}  ,
\end{equation}
which gives
\begin{equation}
\frac{0.3 \, v_{\text{av}}} {R}  \lesssim f \lesssim
\frac{0.72\,  v_{\text{av}}} { R}  .
\end{equation}

\begin{figure}
\begin{center}
 \includegraphics[width=0.9\linewidth]{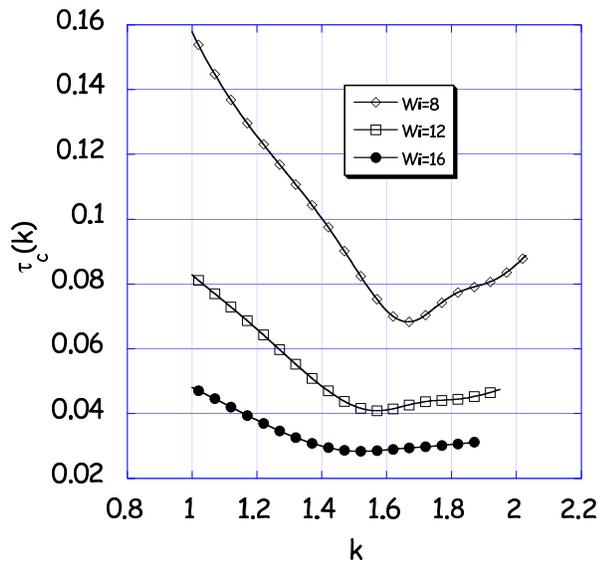}
\end{center}
\caption[]{Plot of the critical amplitude $\tau_{\text{c}}$ as a function of $k$
in the planar case for three different values of ${\sf Wi}$. }\label{f15}
\end{figure}

\begin{figure}
\begin{center}
  \includegraphics[width=0.9\linewidth]{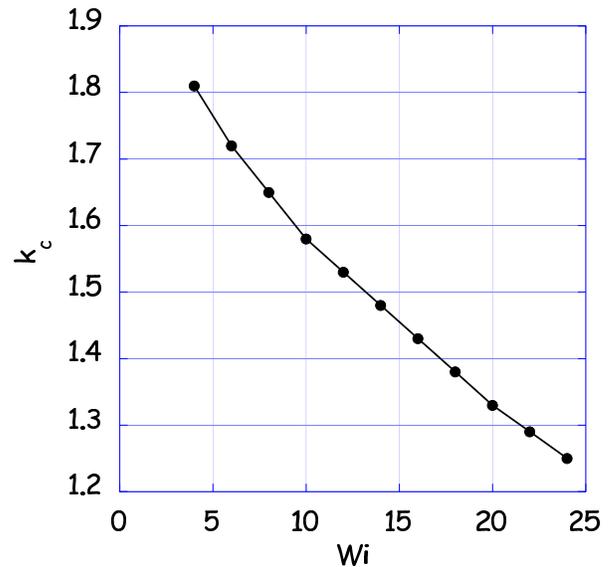}
\end{center}
\caption[]{The value $k_c$ of the wavenumber corresponding to the
minimum of $A_{\text{c}}$ as a function of $k$ for the case of planar
Poiseuille flow. }\label{f16}
\end{figure}

We finally turn to a brief discussion of our results for the case of a
planar slit. The main difference between the cylindrical and planar
case is that in the latter case, the coefficient $\text{Re}\, c_3$ is always
found to be positive.  This is also illustrated by Figs.~\ref{f14}
and \ref{f15}, which show that in the case of the planar slit there is
no finite band of nonlinearly unstable wavenumbers for the critical
shear rate (Fig.~\ref{f14}) and critical shear stress (Fig.~\ref{f15})
as a function of $k$. Note also that in both cases these curves just
have a single minimum, contrary to what we found for the cylindrical
tube. As a result of this, the wavenumber corresponding the 
minimum of the amplitude shear rate critical value $A_{\text{c}}$ now decreases
smoothly with ${\sf Wi}$, see Fig.~\ref{f16}.

The absence of a finite band of wavenumbers and of a value of ${\sf Wi}$
below which $\text{Re}\, c_3 <0$ for all $k$ (as in the cylindrical case),
appears to  have fewer practical implications than one
might expect at first sight. After all, the overall behavior of the minimal
value of the critical amplitudes as a function of ${\sf Wi}$ shows quite the
same behavior as in the case of the cylinder, compare
Figs.~\ref{plaatje4} and \ref{plaatje5}: upon decreasing ${\sf Wi}$, the
critical values rapidly decrease below ${\sf Wi} \approx 5$. As we have
stressed before, our expansion ceases to be valid in this regime where
$A_{\text{c}}$ becomes of order unity. We expect that in reality there is also
a true saddle-node bifurcation point where the branch ends in the case
of planar Poiseuille flow; possibly,
our results indicate that in the planar case the corresponding
critical value ${\sf Wi}_{\text{c}}$ is lower than in the cylindrical
case. However, our results do imply that it is difficult to make a
prediction for the wavenumber near this (presumed) critical value than
in the case of the cylinder. Nevertheless, since the wavenumber of the
mode whose threshold to nonlinear instability is smallest is about 1.8,
it seems reasonable to expect for the dimensional wavelength near threshold
\begin{equation}
{\sf Wi} \approx {\sf Wi}_{\text{c}}: ~~~~~~~~ \Lambda \approx 2 \pi d/1.8
\approx 3.5 \, d
\end{equation}
where the width of the slit equals $2d$.  Like for the case of flow in a pipe, according to
our linear results the phase
velocity is very close to the maximum velocity at the center line between the plates, so the above
result for the most likely wavelength near threshold yields a frequency of about $2.7 \, v_{\text{av}}/d$.
However, since we do not find a finite band of unstable modes above threshold, it is difficult to determine
the range of frequencies expected further above threshold, even though we do expect here too that the
patterns move roughly with the average velocity.

\section{Discussion and outlook} \label{secdisc}

In this article we have shown that for the simplest model of a
polymer fluid with normal stress effects, the UCM model, Poiseuille flow through a
planar channel or cylindrical tube becomes weakly nonlinearly unstable 
for Weissenberg numbers somewhat larger than unity.  Stated differently, since the 
UCM  model {\em only} includes the essential normal stress
effect, we find that the nonlinear flow instability is characterized
{\em by the Weissenberg number only}, and the phenomenon appears to be 
very robust in that almost any more complicated polymer fluid model
that includes normal stress effects will exhibit the same instability
in the same range of Weissenberg numbers.  We presented evidence in
\cite{meulenbroek,bonn1} that this instability yields an intrinsic
route to melt fracture behavior in the absence of other mechanisms such as stick-slip
phenomena.  

\revision{One should also keep in mind that our expansion
is only carried out to lowest order in the nonlinearity, so one may wonder about
the robustness of these results as long as higher order terms in the expansion are 
unknown.  Investigations of these
issues for Couette flow and Poiseuille flow are underway and will be reported in due course.}

The critical  Weissenberg number we find is a factor 10 larger than
the values for which Atalik and Keunings \cite{keunings} observed self-sustained
quasi-periodic oscillations in their numerical simulations of
Poiseuille flow.  The two results are not necessarily inconsistent however: while all of our calculations apply to the UCM model, their
simulations were done for the Oldroyd-B model, with a viscosity ratio of
$10^{-3}$ and Reynolds number
${\sf Re}=0.1$. Moreover, Atalik and Keunings added a stress diffusion term to their equations to improve numerical stability. Further  analysis is clearly necessary to investigate the effects of these
differences --- \revision{a detailed comparison of the simulations with the amplitude expansion
results for one and the same model, is  called for.}

Just above the onset of a well-defined linear (or
``supercritical'') instability threshold, the wavelength of the
pattern is well-defined: upon approach of the threshold, the
wavenumber of the pattern approaches the wavenumber $k_c$ at which the 
instability sets in. Here, however, it is more difficult to draw sharp 
conclusions about the wavelength of the pattern close to onset. First
of all, our expansion can not be fully trusted  quantitatively for
Weissenberg numbers of the order of ${\sf Wi}_{\text{c}}$, as the damping rate is large there, 
not small as is needed in our method. This technical
caveat aside, one should keep in mind that once the instability sets in the 
nonlinear flow behavior which will develop can not be addressed by our
expansion method. Hence it is difficult to  rule out the possibility of nonlinear
interaction terms changing the velocity of the pattern in the die to a
value different from the one we have assumed, namely the average flow
velocity $v_{\text{a}}$. 
 If on the other hand the flow
pattern stabilizes in some weakly nonlinear regime, then one would
expect its wavenumber to be close to the onset value we calculate
and the frequency to be close to the one we have estimated in Eq.~(\ref{transfreq}).
As a rule of thumb,  the wavelength of the undulations of the
extrudate is typically
 about twice the diameter of the die.

In conclusion, our nonlinear analysis
establishes the nonlinear flow instability essentially beyond
reasonable doubt and predicts onset values which are consistent with
those reported experimentally \cite{denn1,pahl,bonn1,bonn2}.
 Moreover, the hypothesis of highly similar subcritical
behavior in Taylor-Couette geometries
\cite{steinberg1,steinberg2} and Poiseuille flow (and hence
possibly melt fracture) is fully confirmed  by our calculations. Indeed, the
similarity between the 
flow field perturbation shown in  Fig.~\ref{velocityplot} and 
the roll-type pattern which Groisman and Steinberg have argued gives rise to the
subcritical nature of the instability \cite{steinberg2} in Taylor-Couette flow is
striking. From this perspective,  the
only difference between the two cases is that the general mechanism \cite{larson,shaqfeh2,pakdel}
that in visco-elastic flows 
the curvature of the streamlines makes the flow {\em  linearly}
unstable  {\em is} operative in Taylor-Couette cells but obviously {\em not } in Poiseuille
flow. Nonlinearly, the two flows appear to be much more closely
connected. From this perspective, the evidence for ``turbulence
without inertia'' \cite{steinbergturb,larsonturb} as a result of these elastic effects is intriguing.

\section{Acknowledgement}
WvS would like to thank Daniel Bonn for stimulating his interest in the
polymer flow problem and Daniel Bonn 
for numerous discussions. Moreover, he is
grateful to the LPS 
of the ENS in Paris for its hospitality and to the French
Minist\`ere des affaires \'etrang\`eres 
and French Minist\`ere de l'\'education nationale,
de la recherche et de la technologie, for a
Descartes-Huygens award which made frequent visits to Paris possible.
 
%\ecols

\begin{appendix}

\section{Explicit expressions for the operator $\mathcal L$ and the nonlinear
term $N$}\label{appendixb}

In this section we give the expressions for
the operators ${\mathcal L}$ and $N$ from the equation
\begin{equation} 
{\mathcal L} V=N(V,V)
\end{equation}
in the planar case and the cylindrical case in dimensional form.

For the planar case we have
\begin{displaymath}
{\mathcal L} =
\left( \begin{array}{ccccc} \de _y & \de _z & 0 & 0 & 0 \\
                              0 & 0 & \de _y \de _z &  \de _z ^2 - \de _y ^2  &  - \de _y \de _z  \\
                              C & 0 & A & 0 & 0 \\ 
                              D & E & B & A & 0 \\
                              F & G & 0 & 2B & A \\
\end{array} \right),
\end{displaymath}
where the operators $A-G$ are defined in the following way:
\begin{eqnarray}
A &=& 1+\lambda \de _t +\lambda \vzy \de _z,
\\
B&=& -\lambda \frac{\de \vzy}{\de y},
\\
C & = & 2(\eta \de _y - \lambda \tau^0 _{yz} (y) \de _z),
\\
D& = & \eta \de _z - \lambda \left( \tau^0 _{zz} (y) \de _z -  \frac{\de
\tau^0_{yz}(y) }{ \de y}\right),
\\
E&=&\eta \de _y,
\\
F & = &\lambda \de _y  \tnuly _{zz},
\\
G&=& -2\lambda (\tau^0 _{zz} (y)\de _z + \tau^0 _{yz} (y) \de _y) + 2\eta \de _z
\end{eqnarray}
and
\begin{eqnarray}  
N_1 & = & 0,   \\
N_1 & = & 0 ,  \\
N_3 & = & \lambda \left(-v_y \frac{\de \tau_{yy}}{\de y} - v_z \frac{\de \tau_{yy}}{\de z}
  \right. \nonumber \\ & & \hspace{2.0cm} \left.  + 2(\frac{\de v_y}{\de y} \tau_{yy} + \frac{\de v_y}{\de z}
    \tau_{yz}) \right) , \\
N_4 & = & \lambda \left(-v_y \frac{\de \tau_{yz}}{\de y} - v_z \frac{\de \tau_{yz}}{\de z}
    + \frac{\de v_z}{\de y} \tau_{yy} \right. \nonumber \\ & &
  \hspace{2.0cm}  \left.  + \frac{\de v_y}{\de z}
    \tau_{zz}  \right), \\
N_5 & = & \lambda \left(-v_y \frac{\de \tau_{zz}}{\de y} - v_z \frac{\de \tau_{zz}}{\de z}
    + 2(\frac{\de v_z}{\de y} \tau_{yz}  \right. \nonumber \\ & &
  \hspace{2.0cm}  \left.  + \frac{\de v_z}{\de z} \tau_{zz})\right).
\end{eqnarray}

For the cylindrical case we get
\begin{equation}
L =
\left( \begin{array}{cccccc} \frac{1}{r}+\de _r & \de _z & 0 & 0 & 0 & 0\\
                              0 & 0 & \de _z(\de _r + \frac{1}{r}) & 0
                              & M & - \de _r \de _z \\
                              B & 0 & A & 0 & 0 & 0\\
                              C & D & E & A & 0 & 0 \\ 
                              F & 0 & 0 & 0 & A & 0 \\
                              G & H & 0 & I & 0 & A \\
\end{array} \right),
\end{equation}
where the operators $A-M$ are defined in the following way:
\begin{eqnarray}
A&=&1+\lambda \de _t +\lambda \vzr \de _z,
\\
B&=&2 \eta \left(\de _r +  \lambda  \frac{\de \vzr}{\de r} \de _z\right),
\\
C&=&\eta \left(\de _z - \lambda \frac{\de^2 \vzr}{\de r^2}-\frac{\lambda}{r} \frac{\de \vzr}{\de r} \right. \nonumber \\ &   & \hspace*{1cm} +\left. 2 \lambda ^2\left(\frac{\de \vzr}{\de r}\right)^2 \de _z\right),
\\
D&=&\eta \de _r,
\\
E&=&-\lambda \frac{\de \vzr}{\de r},
\\
F&=&\frac{2 \eta}{r},
\\
G&=&-4 \eta \lambda ^2  \,\frac{\de \vzr}{\de r} \, \frac{\de^2 \vzr}{\de r^2},
\\
H&=&2 \eta \de _z + \eta \lambda \frac{\de \vzr}{\de r}\left(2 \de _r + 4\lambda \frac{\de \vzr}{\de r} \de _z\right),
\\
I&=&-2 \lambda \frac{\de \vzr}{\de r},
\\
K &=&  -\frac{1}{r} \de _r + \frac{1}{r^2} - \de ^2 _r + \de ^2 _z
\\
M &=& -\frac{1}{r} \de _z 
\end{eqnarray}
and the vector $N$ has components
\begin{eqnarray}   N_1 & = &  0 \\
N_2 & = &  0  \\
N_3 &= & \lambda \left(  -v_r \frac{\de \tau_{rr}}{\de r} - v_z \frac{\de \tau_{rr}}{\de z} + 2\frac{\de v_r}{\de r} \tau _{rr}\right. \nonumber \\ & &  \hspace*{2.0cm} \left. + 2\frac{\de v_r}{\de z} \tau _{rz} \right) \\
N_4 &= & \lambda   \left(
-v_r \frac{\de \tau_{rz}}{\de r} - v_z \frac{\de \tau_{rz}}{\de z}
    + \frac{\de v_z}{\de r} \tau_{zz}\right. \nonumber \\ & &  \hspace*{2.0cm} \left. + \frac{\de v_z}{\de r} \tau_{rz} -\frac{v_r}{r} \tau_{rz}  \right) \\
N_5 &=& \lambda  \left(
-v_r \frac{\de \tau_{\theta \theta}}{\de r} - v_z \frac{\de \tau_{rz}}{\de z}
\right. \nonumber \\ & &  \hspace*{2.0cm} \left.    + 2\frac{v_r}{r} \tau_{\theta \theta} \right)  \\
N_6 &=& \lambda \left(
-v_r \frac{\de \tau_{zz}}{\de r} - v_z \frac{\de \tau_{zz}}{\de z}
  \right. \nonumber \\ & &  \hspace*{1.5cm} \left.  + 2(\frac{\de v_z}{\de r} \tau_{rz} + \frac{\de v_z}{\de z} \tau_{zz}) \right). 
\end{eqnarray}
% \end{appendix}
%\ecols
%\setcounter{appendix}{A}
% \end{multicols}
% \begin{appendix}
% \setcounter{section}{1}
\section{Explicit expressions for the coefficient $\beta_i$}\label{appendixa}

The coefficients of the equation for the streamfunction in the planar case:

\begin{eqnarray}
\beta_0 & := &k^2(k^2+i\,k{\sf S}+2 k^2 {\sf S}^2 \xi^2-6i \,
C(\xi) k{\sf S}^3 \xi^2  \nonumber \\
 & &-i\, C(\xi) 
k{\sf S}  +3C(\xi){\sf S}^2+6i\,C(\xi)^2 k {\sf S}^3 \xi^2 \nonumber \\
 & &+4 C(\xi)^2 {\sf S}^4 k^2 \xi^4+2 C(\xi)^2 k^2 {\sf S}^2 \xi^2
),
\\
\beta_1& := & 2 k^2 {\sf S} \xi (i\,k-2 {\sf S}+4 C(\xi) {\sf S}-2
i\, C(\xi) k {\sf S}^2 \xi^2 \nonumber \\
& & -i\,  C(\xi) k-2 C(\xi)^2 {\sf S}+2i\,C(\xi)^2 k {\sf S}^2 \xi^2),
\\
\beta_2& :=& -k (3i\,{\sf S}+2 k+2 k {\sf S}^2 \xi^2-3i\, C(\xi)
{\sf S} \nonumber \\
 & &-4 C(\xi) k {\sf S}^2 \xi^2+2 C(\xi)^2
 k {\sf S}^2 \xi^2);
\\
\beta_3& :=& 2 i\,k [{\sf S} \xi (-1+C(\xi)),
\\
C(\xi) & :=& 2/(2+i\,k {\sf S} (1-\xi^2-c)),
\end{eqnarray}
where  the complex dimensionless coefficient $c=\omega/k$ is used by
Rothenberger {\em et al} \cite{denn2}.

The coefficients of the streamfunction in the cylindrical case are
\begin{eqnarray}
\beta_0 & := & k^3 (k-4i\, C(\xi) {\sf S}^3 \xi^2+4 k [{\sf S}^4 \xi^4  C(\xi)^2 \nonumber \\
 & &+2 k {\sf S}^2
\xi^2 C (\xi)^2 
+4 i\, C(\xi)^2 {\sf S}^3 \xi^2\nonumber \\
 & &+2 k \xi^2 {\sf S}^2),
\\
\beta_1 & := & (-2 {\sf S}^2 \xi^4 k^2+2 k^2 \xi^2-2i\, k^3 \xi^4
{\sf S} C(\xi) \nonumber \\
 & &-3+4 {\sf S}^2
\xi^4  C(\xi) k^2 
+2 i\, k^3 \xi^4 {\sf S} \nonumber \\
 & &-2 C (\xi)^2 {\sf S}^2 \xi^4 k^2+4i\, k^3 \xi^6  C(\xi)^2 {\sf
S}^3 \nonumber \\
 & &-4i\,
k^3 \xi^6  C(\xi) {\sf S}^3)/(\xi^3),
 \\
\beta_2& :=& -(-3-4 {\sf S}^2 \xi^4  C(\xi) k^2+2 C (\xi)^2 {\sf S}^2 \xi^4 k^2 \nonumber \\
 & &+2 k^2 \xi^2+2
{\sf S}^2 \xi^
4 k^2)/(\xi^2),
\\
\beta_3& :=& 2 (-i\, k \xi^2 {\sf S}-1+i\, {\sf S} \xi^2  C(\xi) k)/\xi,
\\
C(\xi)& :=& 1/(1+i /2 k {\sf S} (1-\xi^2-c)).
\end{eqnarray}

\end{appendix}

%\begin{multicols}{2}

%\end{multicols}

%\ecols


\begin{thebibliography}{99}
\bibitem{notefreq} Note that the band extends to slightly smaller
values at flow rates about 50\% beyond threshold. In this range, the
lower estimate for the frequency might be better replaced by $0.25 v_{\text{av}}/R$.
\end{thebibliography}
\end{document}